\pdfoutput=1
\documentclass[a4paper,13pt]{article}
\usepackage{jheppub} 
\usepackage[utf8]{inputenc}
\usepackage{color,graphicx,slashed,hyperref,ulem}
\hypersetup{
   bookmarks=true,         
   unicode=true,          
   pdftoolbar=true,        
   pdfmenubar=true,        
   pdffitwindow=false,     
   pdfstartview={FitH},    
   pdftitle={My title},    
   pdfauthor={Author},     
   pdfsubject={Subject},   
   pdfcreator={Creator},   
   pdfproducer={Producer}, 
   pdfkeywords={keyword1} {key2} {key3}, 
   pdfnewwindow=true,      
   colorlinks=true,       
   linkcolor=blue,          
   citecolor=magenta,        
   filecolor=magenta,      
   urlcolor=cyan           
}

\definecolor{red}{rgb}{1.0, 0, 0}
\definecolor{orange}{rgb}{1,0.,1}
\newcommand{\be}{\begin{equation}}
\newcommand{\ee}{\end{equation}}
\def\bsp#1\esp{\begin{split}#1\end{split}}
\newcommand{\ie}{\textit{i.e.}}
\newcommand{\eg}{\textit{e.g.}}

\begin{document}
\date{\today}

\title{Investigating light NMSSM pseudoscalar states with boosted ditau tagging}

\author[a]{Eric Conte,}
\author[b,c]{Benjamin Fuks,}
\author[d,e]{Jun Guo,}
\author[f]{Jinmian Li,}
\author[f]{Anthony G.\ Williams}

\affiliation[a]{Groupe de Recherche de Physique des Hautes \'Energies (GRPHE),
   Universit\'e de Haute-Alsace, IUT Colmar, 34 rue du Grillenbreit BP 50568,
   68008 Colmar Cedex, France}
\affiliation[b]{Sorbonne Universit\'es, UPMC Univ.~Paris 06,
  UMR 7589, LPTHE, F-75005, Paris, France}
\affiliation[c]{CNRS, UMR 7589, LPTHE, F-75005, Paris, France}
\affiliation[d]{State Key Laboratory of Theoretical Physics, Institute of Theoretical Physics,
Chinese Academy of Sciences, Beijing 100190, P. R. China}
\affiliation[e]{Institut Pluridisciplinaire Hubert Curien/D\'epartement Recherches
 Subatomiques, Universit\'e de Strasbourg/CNRS-IN2P3,
 23 rue du Loess, F-67037 Strasbourg, France}
\affiliation[f]{ARC Centre of Excellence for Particle Physics at the Terascale and CSSM, Department of Physics, University of Adelaide, Adelaide, SA 5005, Australia}

\emailAdd{eric.conte@iphc.cnrs.fr}
\emailAdd{fuks@lpthe.jussieu.fr}
\emailAdd{hustgj@itp.ac.cn}
\emailAdd{phyljm@gmail.com}
\emailAdd{anthony.williams@adelaide.edu.au}


\abstract{
We study a class of realizations of the Next-to-Minimal Supersymmetric Standard
Model that is motivated by dark matter and Higgs data, and in which the lightest
pseudoscalar Higgs boson mass is smaller than twice the bottom quark mass and
greater than twice the tau lepton mass. In such scenarios, the lightest
pseudoscalar Higgs boson can be copiously produced at the LHC from the decay of
heavier superpartners and will dominantly further
decay into a pair of tau leptons that
is generally boosted. We make use of a boosted object tagging technique designed
to tag such a ditau jet, and estimate the sensitivity of the LHC to the
considered supersymmetric scenarios with 20 to 50~fb$^{-1}$ of proton-proton
collisions at a center-of-mass energy of 13~TeV.
}

\maketitle
\flushbottom

\section{Introduction}
\label{sec:intro}
The Standard Model of particle physics has been proven an extremely successful
theory of nature, but it leaves many questions unanswered. It is consequently
widely acknowledged as an effective theory obtained from a more fundamental
theoretical context still to be observed. Supersymmetric extensions of the
Standard Model represent one of the most popular options for new physics and are
motivated by the unification of gauge and space-time symmetries. In addition,
they resolve the hierarchy problem inherent to the Standard Model, feature the
unification of the gauge couplings at high energy scales and naturally provide
an explanation to the presence of dark matter in the universe. By construction,
the Higgs sector of a supersymmetric theory is extended with respect to the
Standard Model case and contains at least two weak doublets of Higgs superfields
(traditionally noted $H_u$ and $H_d$) so that masses for both the up-type and
down-type fermions could be generated. Considering the minimal supersymmetric
extension of the Standard Model, the so-called Minimal Supersymmetric Standard
Model (MSSM)~\cite{Nilles:1983ge,Haber:1984rc}, only the $H_u$ and $H_d$ Higgs
supermultiplets are included and the superpotential contains a supersymmetric
mass term for these superfields $\mu H_u H_d$. While the dimensionful parameter
$\mu$ should in principle be of the order of the only natural scale of the
theory that is either the Planck or the gauge-coupling unification
scale, a working spontaneous breaking of the electroweak symmetry demands this
parameter to be in the ball park of a few hundreds of GeV. This puzzle is called
the `$\mu$-problem' of the MSSM~\cite{Kim:1983dt}. On different lines, the
discovery of a
scalar field with a mass of about 125~GeV and that resembles the Standard Model
Higgs boson~\cite{Aad:2012tfa,Chatrchyan:2012xdj} implies either the existence
of heavy top squarks or large top squark mixing, which raises questions about
the naturalness of the MSSM.

All these issues can be solved elegantly in the framework of the Next-to-Minimal
Supersymmetric Standard Model (NMSSM)~\cite{Ellwanger:2009dp}, where the model
includes an additional superfield $S$ that is singlet under the Standard Model
gauge group. As a result, the Higgs sector of the model features three neutral
scalar states, two neutral pseudoscalar states and one charged state, as well as
one singlino (the fermionic component of $S$) and two higgsinos (the fermionic
components of $H_u$ and $H_d$) fermions that will mix with the gauginos to form
five neutralinos. This enriched particle content yields a phenomenology that
could be largely different from the MSSM case and that could even
accomodate~\cite{Ellwanger:2016qax} the tantalizing hints of an excess of
diphoton events observed in LHC data at a center-of-mass energy of
13~TeV~\cite{ATLAS-CONF-2015-081,CMS:2015dxe}.

In order to satisfy the stringent constraints on the Standard Model Higgs boson
properties derived from LHC measurements~\cite{Khachatryan:2014jba,Aad:2015gba},
and in particular those that are put on the Higgs exotic decay modes,
phenomenologically viable NMSSM scenarios have to contain a
Standard-Model-like Higgs boson with a very small singlet component.
Consequently, one given scalar state and one given pseudoscalar state have to be
almost purely singlet, so that they couple to the Standard Model only through
their small mixing with the $H_u$ and $H_d$ fields. Furthermore, these singlet
fields are weakly constrained by current experimental data and can hence be as
light as a few GeV. This setup with two light singlet-like bosons is
further motivated by the Peccei-Quinn symmetry limit of the NMSSM where one
imposes that the model Lagrangian is invariant under a Peccei-Quinn-like
symmetry. This indeed not only allows the NMSSM to solve the
strong $CP$-problem~\cite{Miller:2003ay}, but also yields a very light
pseudoscalar singlet state $A_1$. Such a prediction has spurred an intense
phenomenological activity over the last years~\cite{Dermisek:2005ar,%
Dermisek:2006wr,Ellwanger:2005uu,Djouadi:2008uw,Cao:2013gba,Bomark:2014gya,%
Bomark:2015fga,Potter:2015wsa}, with a particular focus on processes where light
$A_1$ pairs are produced from the cascade decays of heavier Higgs bosons.
Pioneering works have investigated final state systems made of four jets issued
from the fragmentation of
$b$-quarks \mbox{$H\to A_1 A_1\to b\bar{b}b\bar{b}$}~\cite{Almarashi:2011te},
four leptonically or hadronically decaying tau leptons \mbox{$H\to A_1 A_1\to
\tau^+ \tau^- \tau^+ \tau^-$}~\cite{Forshaw:2007ra,Belyaev:2008gj,%
Cerdeno:2013cz}, four muons \mbox{$H\to A_1 A_1\to \mu^+\mu^-\mu^+\mu^-$}%
~\cite{Belyaev:2010ka} or of one pair of muons and one pair of tau leptons
\mbox{$H\to A_1 A_1\to \mu^+ \mu^- \tau^+ \tau^-$}~\cite{Lisanti:2009uy}. It has
been moreover shown that the discovery of such decay channels would consist of a
no-lose theorem for a direct evidence of the NMSSM~\cite{Ellwanger:2003jt,%
Ellwanger:2013ova}. This has consequently opened the path for dedicated NMSSM
searches in LHC collision data at a center-of-mass energy of 8~TeV~\cite{%
Khachatryan:2015nba,Aad:2015oqa,Khachatryan:2015wka}. Upper limits on the
production cross sections related to the four taus, two taus and two muons, and
four muons decay modes of the heavy Higgs boson of 4.5--10.3~pb, 0.72--2.33~pb
and of about 1~fb have been respectively derived.

In this work, we study Higgs data constraints on the NMSSM and
show that phenomenologically viable scenarios feature configurations in which
the lightest pseudoscalar $A_1$ is dominantly produced from the decays of
neutralino states~\cite{Cheung:2008rh,Cerdeno:2013qta,Han:2015zba}. The LHC
constraints on these scenarios are still both rather weak and very model
dependent. In particular, $\tilde{\chi}^0_i \to A_1 \tilde{\chi}^0_j$ decays
often lead to the production of boosted $A_1$ particles that are difficult to
detect due to the collimation of their decay products into a single object,
regardless of the mass splitting between the two neutralinos $\tilde{\chi}^0_i$
and $\tilde{\chi}^0_j$. We explore in Section~\ref{sec:nmssm} the parameter
space of the NMSSM and investigate specific scenarios compatible with the
above-mentioned Higgs requirements
and featuring a light pseudoscalar state $A_1$, with a focus on cases where its
mass is of at least twice as large as the tau lepton mass and smaller than twice
the mass of the $b$-quark. In our process for constructing such scenarios, we
additonally impose dark matter considerations on the lightest
supersymmetric partner. We then
investigate, in Section~\ref{sec:simu}, the
sensitivity of the current LHC run at 13~TeV to such scenarios and show that
they could be detected through the analysis of a signature comprised of a single
lepton, a ditau-tagged jet and missing transverse energy. To this aim, we make
use of a ditau tagging technique that has been developed in the context of
the Higgs~\cite{Englert:2011iz,Papaefstathiou:2014oja,Katz:2010iq} and that
we have supplemented to a multivariate analysis dedicated to the tagging of the
signal. Our conclusions are summarized in Section~\ref{sec:conl}.

\section{Light scalar and pseudoscalar Higgs bosons in the NMSSM}
\label{sec:nmssm}

\subsection{Theoretical framework}
The NMSSM is constructed by augmenting the MSSM superfield content by one
superfield $S$ that is a singlet under $SU(3)_c\times SU(2)_L\times U(1)_Y$.
After the breaking of supersymmetry (and the consequent breaking of the
electroweak symmetry), the scalar component of $S$ mixes with the Higgs
degrees of freedom, whilst the fermionic component of $S$, dubbed the
singlino, mixes with the two remaining higgsino states and the gauginos. As for
any softly broken supersymmetric theory, the NMSSM is specified by its
superpotential and its supersymmetry-breaking Lagrangian. The superpotential
reads
\be
  W_{\rm NMSSM} =
    - L H_d \mathbf{y_e} E
    - Q H_d \mathbf{y_d} D
    + Q H_u \mathbf{y_u} U
    + \lambda S H_u H_d
    + \frac13 \kappa S^3\ ,
\label{eq:wnmssm}\ee
where all indices are omitted for brevity, where $Q$ and $L$ denote the weak
doublets of quark and lepton superfields and where $U$, $D$ and $E$ are the
up-type quark, down-type quark and lepton weak-singlet superfields,
respectively. In addition, we have
introduced the $3\times 3$ Yukawa matrices $\mathbf{y}$ and the $\lambda$
and $\kappa$ parameters that drive the couplings of $S$. In particular, once
the scalar component $s$ of the singlet superfield gets a vacuum expectation
value $\langle s\rangle = v_s/\sqrt{2}$, an effective $\mu$-term is generated,
\be
  \mu_{\rm eff} = \frac{1}{\sqrt{2}} \lambda v_s \ ,
\ee
which solves the MSSM $\mu$-problem. In the expression of Eq.~\eqref{eq:wnmssm},
we have imposed that the superpotential satisfies a $\mathbb{Z}_3$ symmetry so
that any dimensionful term allowed by the gauge symmetry is forbidden.

The soft supersymmetry breaking Lagrangian contains mass terms for all scalar
(${\bf m^2_{\tilde Q}}$, ${\bf m^2_{\tilde U}}$, ${\bf m^2_{\tilde D}}$,
${\bf m^2_{\tilde L}}$, ${\bf m^2_{\tilde E}}$, $m_{H_u}^2$, $m_{H_d}^2$ and
$m_s^2$) and gaugino ($M_1$, $M_2$ and $M_3$) fields, as well as trilinear
interaction terms (${\bf A^u}$, ${\bf A^d}$, ${\bf A^e}$, $A_\lambda$ and
$A_\kappa$) sharing the form of the superpotential,
\be \bsp 
  \mathcal{L}_{\text{soft}} =
   &\ - \frac12 \Big[
     M_1\ \tilde B \tilde B +
     M_2\ \tilde W \tilde W +
     M_3\ \tilde g \tilde g +
     {\rm h.c.} \Big]\\
   &\ - {\bf m^2_{\tilde Q}}\  \tilde q^\dag \tilde q
      - {\bf m^2_{\tilde U}}\  \tilde u^\dag \tilde u
      - {\bf m^2_{\tilde D}}\  \tilde d^\dag \tilde d
      - {\bf m^2_{\tilde L}}\  \tilde \ell^\dag \tilde \ell
      - {\bf m^2_{\tilde E}}\  \tilde e^\dag \tilde e
      - m_{H_u}^2\ h_u^\dag h_u
      - m_{H_d}^2\ h_d^\dag h_d
      - m_s^2\ s^\dag s\\
   &\ + \Big[ - {\bf y^u}{\bf A^u} \ \tilde u^\dag \tilde q h_u
              + {\bf y^d}{\bf A^d} \ \tilde d^\dag \tilde q h_d
              + {\bf y^e}{\bf A^e} \ \tilde e^\dag \tilde \ell h_d
              - \lambda A_\lambda h_u h_d s
              - \frac13 \kappa A_\kappa s^3 + {\rm h.c.} \Big]\ ,
\esp\label{eq:lsoft} \ee
where $q$, $\ell$, $u$, $d$, $e$, $h_u$, $h_d$ and $s$ denote the scalar
components of the $Q$, $L$, $U$, $D$, $E$, $H_u$, $H_d$ and $S$ superfields,
respectively, and $\tilde B$, $\tilde W$ and $\tilde g$ the gauginos associated
with the $U(1)_Y$, $SU(2)_L$ and $SU(3)_C$ gauge groups. All indices are again
understood for clarity.

In order to reduce the number of free parameters, we assume that all parameters
related to the (s)fermion sector are flavor-conserving and universal at the
grand unification scale. Introducing the common scalar mass $m_0$, the common
gaugino mass $m_{1/2}$ and the common trilinear coupling $A_0$, we have
\be
   {\bf m^2_{\tilde Q}} = {\bf m^2_{\tilde U}} = {\bf m^2_{\tilde D}} =
     {\bf m^2_{\tilde L}} = {\bf m^2_{\tilde E}} = 
     m_0^2 \ \mathbf{1}_{3\times 3} \ ,~~
   M_1 = M_2 = M_3 = m_{1/2} \ ,~~
   {\bf A^u} = {\bf A^d} = {\bf A^e} = A_0 \ \mathbf{1}_{3\times 3} \ ,
\ee
where $\mathbf{1}_{3\times 3}$ stands for the identity matrix in flavor space.

In this framework, the Higgs sector is defined by the soft parameters
$A_\lambda$ and
$A_\kappa$ that we fix at the grand unification scale, and by the $\lambda$,
$\kappa$, $\tan\beta$ and $\mu_{\rm eff}$ parameters that are provided at the
electroweak scale, $\tan\beta$ being the ratio of the vacuum expectation values
of the neutral components of the two Higgs doublets $h_u$ and $h_d$.

\subsection{Exploration of the NMSSM parameter space}
\begin{table}
\footnotesize
\center
  \setlength{\tabcolsep}{2.5mm}
  \renewcommand{\arraystretch}{1.4}
  \begin{tabular}{c|c|c|c|c}
  $m_0$ & $m_{1/2}$ & $A_0$ & $A_\lambda$ & $A_\kappa$\\\hline
  [400, 2000]~GeV & [1000, 2000]~GeV & [-5000, -1000]~GeV & [-500, 500]~GeV &
    [0, 300]~GeV\\
  \end{tabular}\\[.3cm]

  \begin{tabular}{c|c|c|c}
    $\lambda$ & $\kappa$ & $\tan \beta$ & $\mu$\\ \hline
    [0.2, 0.5] & [0.01, 0.2] & [1.5, 15] & [100, 350]~GeV \\
  \end{tabular}
  \caption{Parameterization of the NMSSM parameter space explored in this work.
    We indicate the ranges in which the different parameters have been allowed
    to vary. In the first table, the parameters are provided at the grand
    unification scale while in the second table, they are given at the
    electroweak scale.}
  \label{tab:range}
\end{table}
In the previous section, we have defined a parameterization of the NMSSM in
terms of nine free parameters,
\be
  m_0,\qquad   m_{1/2},\qquad   A_0,\qquad  A_\lambda,\qquad   A_\kappa,\qquad
  \lambda,\qquad    \kappa, \qquad
  \tan\beta \quad \text{and}\quad \mu_{\rm eff},
\label{eq:susyprm}\ee
the first five parameters being defined at the grand unification scale and the
last four parameters being defined at the electroweak scale. We supplement to
these the parameters related to the Standard Model sector,
\be\bsp
 &
  \alpha_s(m_Z) = 0.1172,\qquad
  G_F = 1.16639\ 10^{-5}~{\rm GeV}^{-2},\qquad
  \alpha(m_Z) = 1/127.92,\qquad
  m_Z = 91.187~\text{GeV},\\
 &\hspace*{2.5cm}
  m_t^{\rm pole} = 173.1~\text{GeV},\qquad
  m_b(m_b) = 4.214~\text{GeV},\qquad
  m_\tau = 1.777~\text{GeV}.
\esp\ee
The QCD interaction strength is computed from the value of the strong coupling
constant at the $Z$-pole $\alpha_s(m_Z)$ and the three independent electroweak
inputs, whose values are taken from the Particle Data Group
review~\cite{Agashe:2014kda}, are chosen to be the Fermi constant $G_F$, the
$Z$-boson mass $m_Z$ and the electromagnetic coupling evaluated at the $Z$-pole
$\alpha(m_Z)$. Finally, the third generation fermion sector is defined by the
pole mass of the top quark $m_t^{\rm pole}$, the running $\overline{\rm MS}$
mass of the bottom quark $m_b(m_b)$ evaluated at the $m_b$ scale and the tau
mass $m_\tau$, all other fermion masses being neglected. All the couplings and
masses appearing in the NMSSM Lagrangian can be subsequently numerically
calculated, using in particular the relations determined by the minimization of
the scalar potential.

For our exploration of the NMSSM parameter space, we use the {\sc NmssmTools}
package~\cite{Ellwanger:2006rn,Das:2011dg,Muhlleitner:2003vg} and perform a scan
over the parameters given in Eq.~\eqref{eq:susyprm}. The ranges in which the
parameters are allowed to vary are given in Table~\ref{tab:range}, and
for scanned each
point, we impose a set of constraints that allows us to accept it or reject it.
Additionally to theoretical considerations such as obtaining a physical spectrum
that does not exhibit any tachyonic state or preventing the appearance of a
Landau pole below the grand unification scale, we impose limits on the Higgs
sector and on the supersymmetric particles derived from collider searches at
LEP, at the Tevatron and at the LHC. We moreover verify the consistency of the
selected points with dark matter data.

In order to accomodate a Higgs boson with properties close to those expected in
the case of the Standard Model, the mixing between the singlet state $s$ and the
two Higgs doublets $h_u$ and $h_d$ has to be small. In this case, the tree-level
masses of the $CP$-even and $CP$-odd singlet-like Higgs bosons are mostly given
by the corresponding entries in the $3 \times 3$ scalar and pseudoscalar squared
mass matrices ${\cal M}^2_S$ and ${\cal M}^2_P$,
\begin{align}
  ({\cal M}_S^2)_{33}  &=
       \lambda A_\lambda \frac{ v^2 \sin 2 \beta}{2 v_s}
     + \frac{\kappa v_s}{\sqrt{2}} \Big(A_\kappa+2\sqrt{2}\kappa v_s\Big)~,~
    \label{eq:mh1} \\
  ({\cal M}_P^2)_{33}  &=
       \lambda A_\lambda \frac{ v^2 \sin 2 \beta}{2 v_s}
     + \frac{\kappa v_s}{\sqrt{2}}
         \Big(\frac{\sqrt{2} \lambda v^2\sin 2\beta}{v_s}  - 3 A_\kappa\Big)
   \label{eq:ma1}~,~
\end{align}
where the vacuum expectation value $v$ is defined, at the tree level, by
\be
  v^2 =  \frac{4 m_Z^2}{g_1^2 + g_2^2}\ ,
\ee
$g_1$ and $g_2$ denoting the hypercharge and weak coupling constants.
In this work, we focus on the phenomenology of scenarios in
which the lightest
pseudoscalar state $A_1$ is mostly singlet-like, so that
$m^2_{A_1}\approx ({\cal M}^2_P)_{33}$, and where its mass is heavier than twice
the tau lepton mass and lighter than twice the bottom quark mass,
$m_{A_1} \in [2 m_\tau, 2 m_b]$. As a consequence, a cancellation between the
different terms of Eq.~\eqref{eq:ma1} should be in place. Since the natural size
of each term is of about the electroweak scale squared
$\mathcal{O}(10^4)$~GeV$^2$, the mass of the lightest singlet-like scalar state,
that is approximately given by Eq.~\eqref{eq:mh1} in the absence of a too large
singlet-doublet mixing, will be smaller than 125~GeV. It turns out that the
second scalar Higgs particle $H_2$ has to be identified with the boson
discovered during the first run of the LHC. Its mass $m_{H_2}$ can moreover
be generally compatible with the observed value of 125~GeV, the presence of
the lighter singlet-like $H_1$ state helping to increase it through mixing
effects~\cite{Kang:2012sy}. We thus impose in our scan that $m_{H_2}$ lies in
the $[122.1, 128.1]$~GeV mass window, such a large range allowing us to account
for the various sources of uncertainties both on the experimental result and on
the theoretical prediction.

Beside its mass, the first run of the LHC has allowed one to measure a lot of
properties of the Higgs boson with high precision. One specific ensemble of such
measurements consists of the so-called Higgs signal strengths $\mu_{X,Y}$ that
are defined as the ratios of predicted rates in a particular production channel
$X$ and decay mode $Y$ of the Higgs boson in a given new physics theory (being
here the NMSSM) to the Standard Model expectation,
\be
  \mu_{X,Y} = \epsilon_{X,Y}\ \frac{\sigma_X}{\sigma^{\rm SM}_X}\
      \frac{{\rm BR}(h\to Y)}{{\rm BR}^{\rm SM}(h\to Y)}\ .
\ee
The $\epsilon_{X,Y}$ factor in the above expression is related to the acceptance
and efficiency of the analyses under consideration, and could be different in a
new physics context and in the Standard Model. Such differences are however
usually assumed to be mild and to largely cancel in the ratio. The signal
strengths moreover depend on the Higgs production cross sections in the
$X$ channel ($\sigma_X^{\rm SM}$ and $\sigma_X$) and branching ratios associated
with the $h\to Y$ decay mode (${\rm BR}(h\to Y)$ and
${\rm BR}^{\rm SM}(h\to Y)$) both in the Standard Model and
in the NMSSM. Making use of the NMSSM signal strengths predicted by {\sc
NmssmTools} and the experimental measurements at the LHC~\cite{Aad:2015gba,%
Khachatryan:2014jba}, we construct a $\chi^2$ quantity~\cite{Bernon:2014vta}
\begin{align}
\chi_Y^2 = & a_Y (\mu_{\text{ggF},Y} - \hat{\mu}_{\text{ggF},Y})^2 + 2 b_Y (\mu_{\text{ggF},Y} - \hat{\mu}_{\text{ggF},Y}) (\mu_{\text{VBF+VH},Y} - \hat{\mu}_{\text{VBF+VH},Y}) \nonumber \\
  &+ c_Y (\mu_{\text{VBF+VH},Y} - \hat{\mu}_{\text{VBF+VH},Y})^2~,~
\end{align}
where the `ggF', `VBF' and 'VH' indices respectively refer to the gluon-fusion,
vector-boson fusion and Higgs-Strahlung Higgs production mechanisms. The
$\hat{\mu}_{X,Y}$, $a_Y$, $b_Y$ and $c_Y$ parameters resulting from the fit of
the LHC Higgs measurements are taken from the {\sc Lilith}-1.1.3 program using
the 15.09 database~\cite{Bernon:2015hsa}, and we require $\chi_Y^2 < 6.18$ so
that our NMSSM predictions are compatible at the $2\sigma$ level with data.

\begin{figure}
  \centering
  \includegraphics[width=0.48\textwidth]{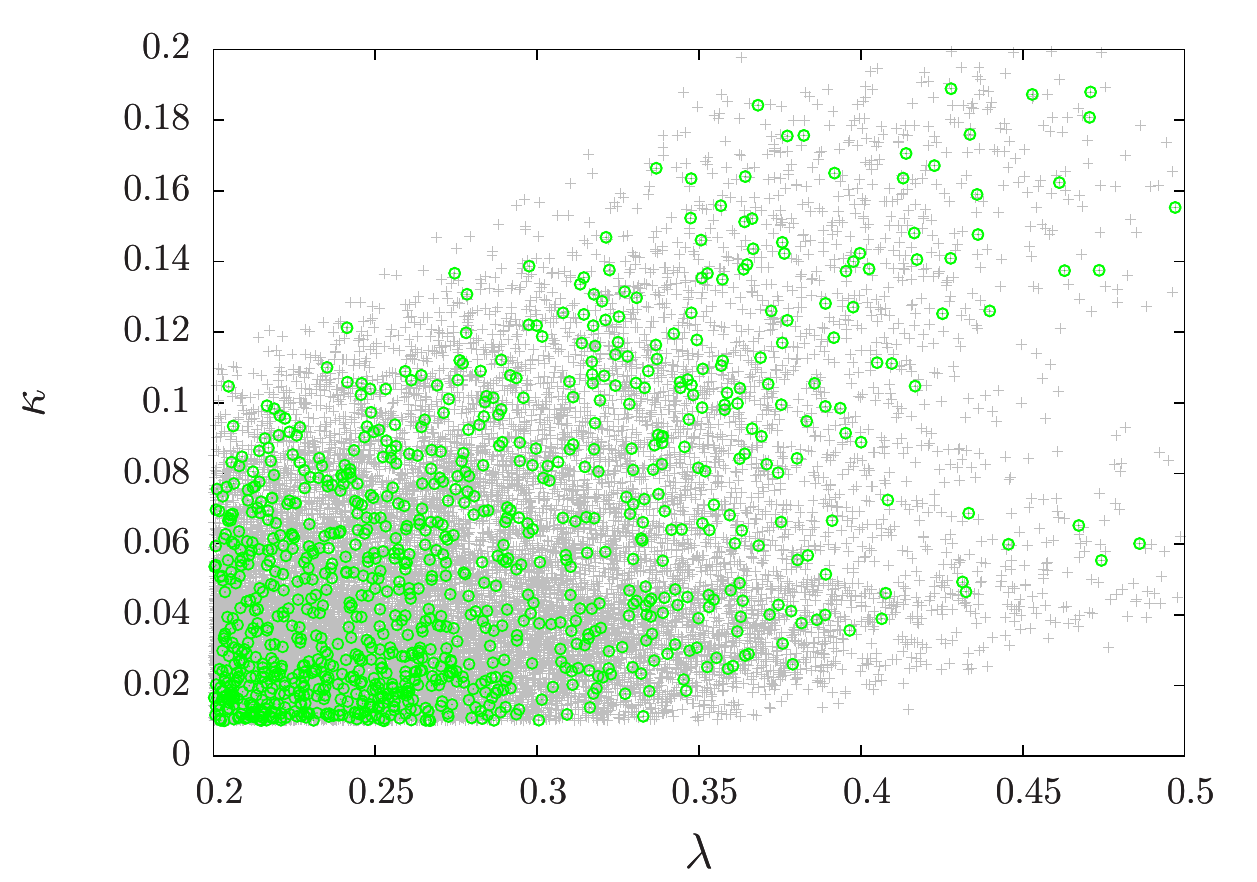}
  \includegraphics[width=0.48\textwidth]{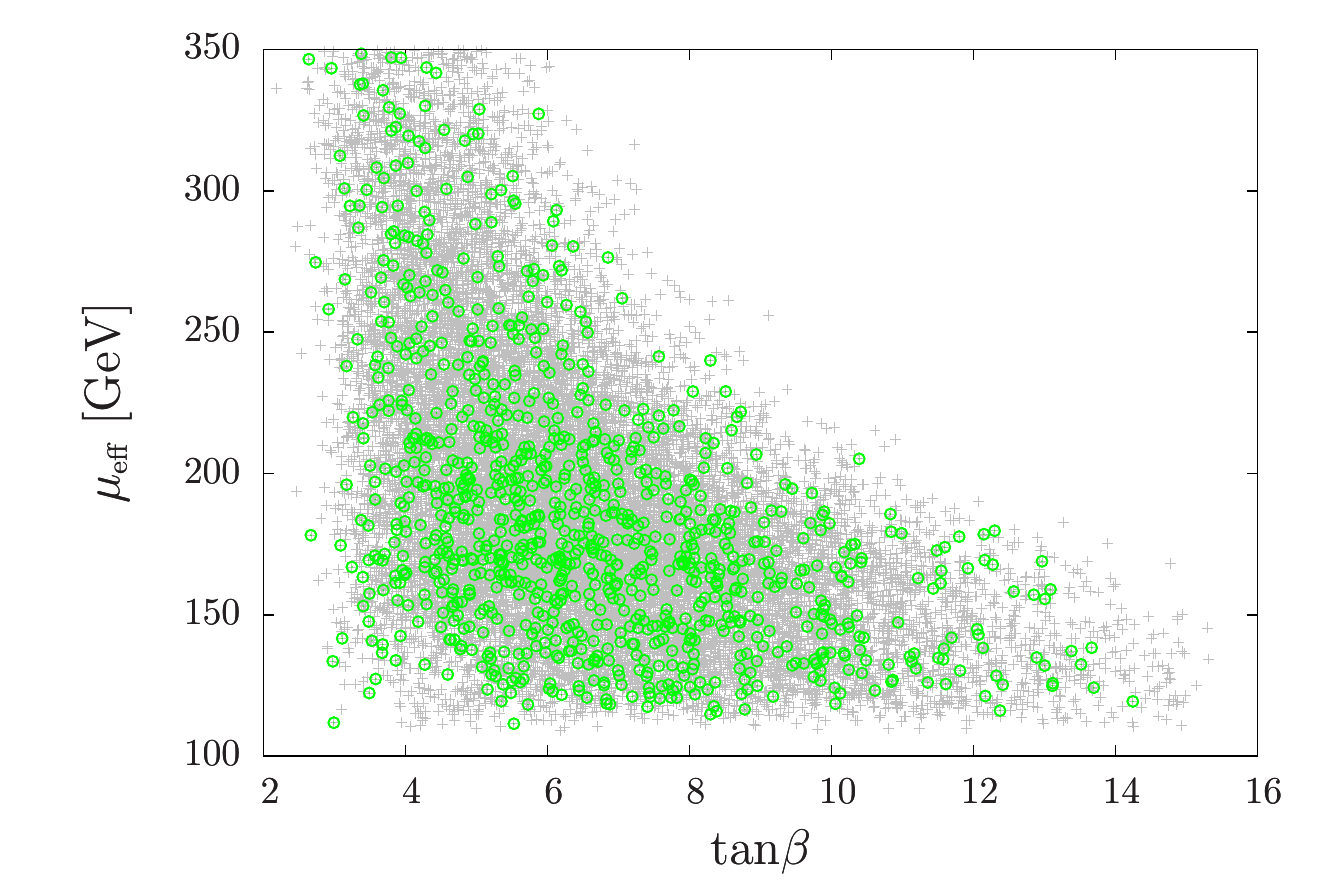} \\
  \includegraphics[width=0.48\textwidth]{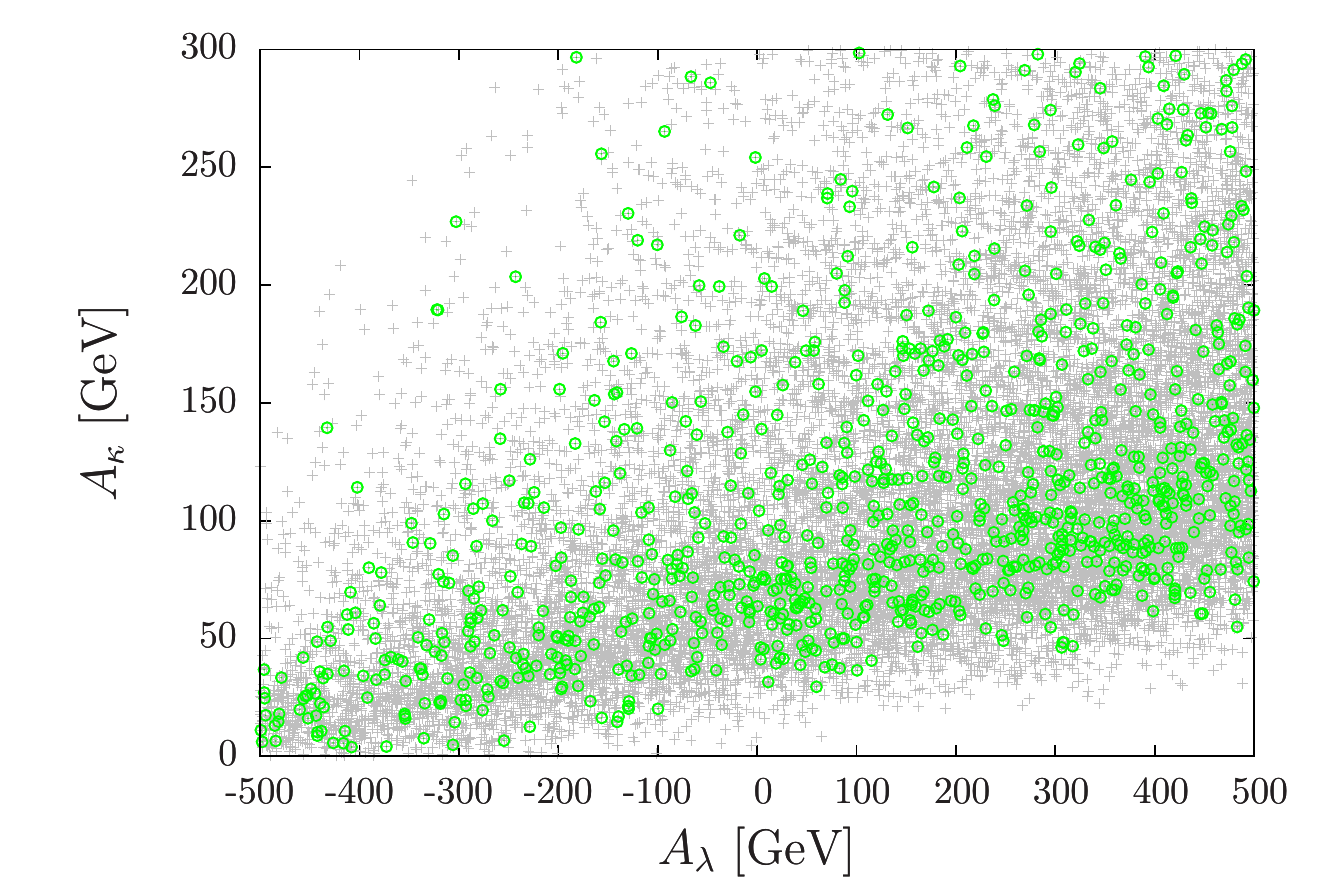}
  \includegraphics[width=0.48\textwidth]{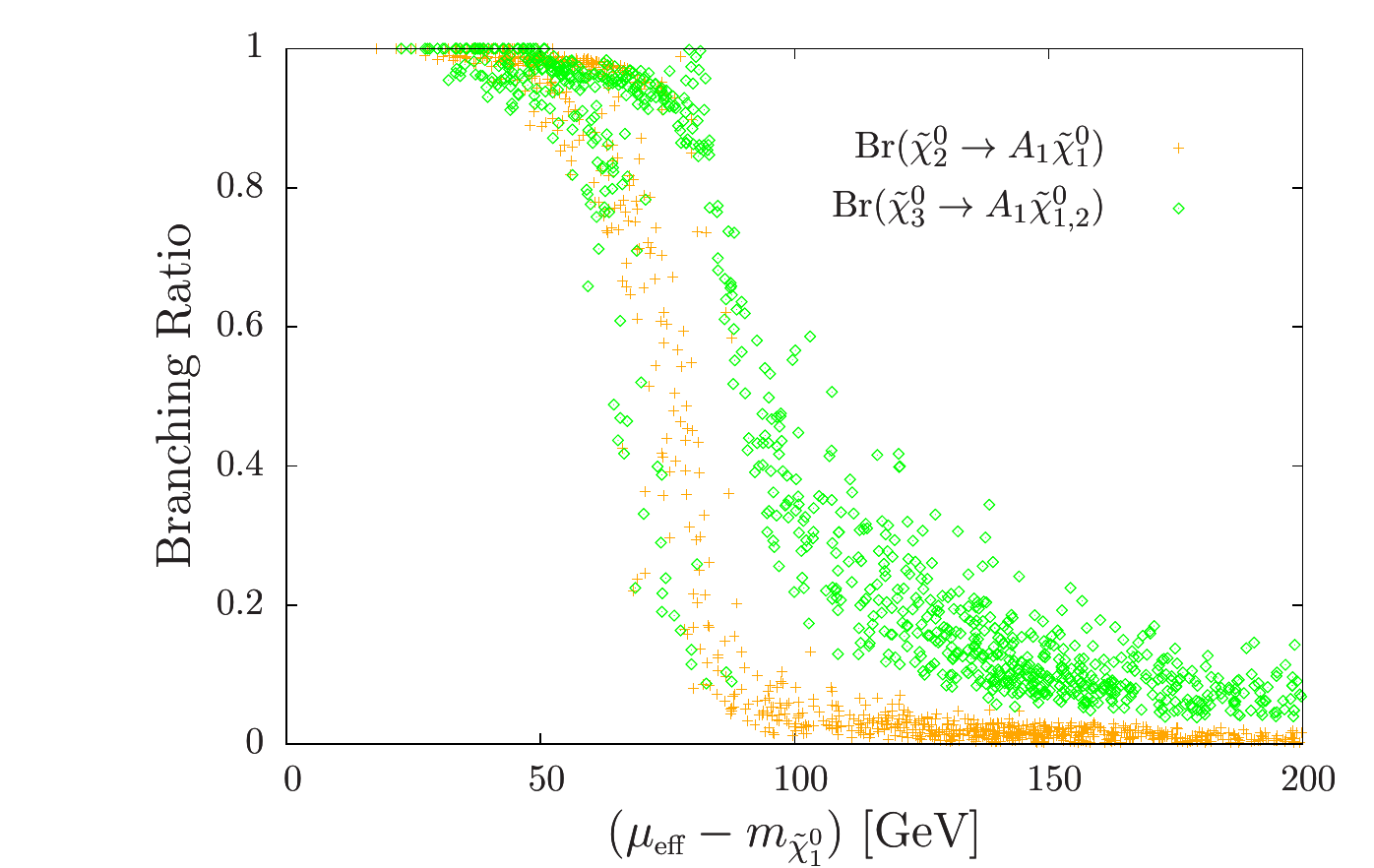} 
  \caption{\label{fig:scan} Benchmark NMSSM scenarios selected in our scanning
   procedure of the NMSSM parameter space, presented in the $(\lambda, \kappa)$
   (upper left panel),
   $(\tan\beta, \mu_{\rm eff})$ (upper right panel) and $(A_\lambda, A_\kappa)$
   (lower left panel) planes. All models for which a physical NMSSM spectrum at
   the electroweak scale is found are represented, under the condition that the
   second lightest scalar state $H_2$ is Standard-Model-like. We distinguish the
   cases where $m_{A_1}<30$~GeV (gray) and where $2 m_\tau < m_{A_1} < 2 m_b$
   (green). In the lower right panel, we depict the dependence of the second and
   third neutralino branching fractions into a pseudoscalar $A_1$ state on the
   compressivity of the spectrum defined as $\mu_{\rm eff}-m_{\tilde\chi_1^0}$,
   for all points satisfying $2 m_\tau < m_{A_1} < 2 m_b$.}
\end{figure}
All the points satisfying the requirements introduced so far are represented in
Figure~\ref{fig:scan} as
a function of the parameters defining the NMSSM Higgs sector. More precisely, we
show results in the $(\lambda, \kappa)$, $(\tan\beta, \mu_{\rm eff})$ and
$(A_\lambda, A_\kappa)$ planes in which correlations can be observed. Each
represented point refers to a particle spectrum free from any tachyonic state
and in which the second lightest Higgs boson $H_2$ has a mass comprised in the
[122.1, 128.1]~GeV range and yields a good $\chi^2$ fit of the measured
signal strengths. We indicate the scenarios for which the pseudoscalar mass is
smaller than 30~GeV by gray points, while the green points correspond to cases
in which $2 m_\tau < m_{A_1} < 2 m_b$. It turns out that small $\lambda$ and
$\kappa$ values are generally favored, which results in a light $A_1$ state as
given by
Eq.~\eqref{eq:ma1}. On different lines, the $\tan\beta$ and $\mu_{\rm eff}$
parameters are strongly correlated, and a large (small) value of $\tan\beta$
leads to a small (large) $\mu_{\rm eff}$ value. This feature originates from
imposing that the $H_2$ scalar boson is Standard-Model-like.
Finally, an approximately linear correlation between the $A_\lambda$ and
$A_\kappa$ parameters is shown in the lower left panel of the figure. This
arises from the singlet-doublet mixing that is more conveniently
assessed when the scalar components of the Higgs fields are rotated to the
so-called Higgs basis $(H_{\rm SM}, H', S')$ by means of an appropriate $U(2)$
transformation~\cite{Ginzburg:2004vp,Davidson:2005cw,Haber:2006ue}. This basis
choice has the advantage that only one of the non-singlet fields, $H_{\rm SM}$,
acquires a vacuum expectation value $v$ and features Standard-Model-like
interactions with the Standard Model fermions and gauge bosons. The second
non-singlet field $H'$ has thus a vanishing vacuum expectation value and the
third basis element $S'$ remains a pure singlet state. In the Higgs basis, the
size of the singlet-doublet mixing is given by the \mbox{$H_{\rm SM}-S'$}
element of the squared mass matrix~${\cal M}^{\prime 2}_S$~\cite{Miller:2003ay,%
Ellwanger:2015uaz},
\be
  ({\cal M}_S^{\prime2})_{13} = \frac{\lambda v}{\sqrt{2}}
    \Big(2 \mu_{\rm eff}-(A_\lambda + \sqrt{2}\kappa v_s) \sin 2\beta\Big)~.~
\ee
In order for $H_2$ to be Standard-Model-like and thus mostly equivalent to
$H_{\rm SM}$ in the Higgs basis, $({\cal M}_S^{\prime2})_{13}$ must be small,
which leads to
\be
  A_\lambda \sim \frac{2 \mu_{\rm eff}}{\sin 2 \beta} -
     \frac{2 \kappa \mu_{\rm eff}}{\lambda}\ .
\label{eq:alconstraint}\ee
However, in the NMSSM parameter space region of interest, we have
$\sin 2\beta\ll\lambda/\kappa$. Consequently, only the first term of the
right-hand side of Eq.~\eqref{eq:alconstraint} matters. Moreover,
$\mu_{\rm eff}$ and $\tan\beta$ are correlated, as shown in the upper right
panel of Figure~\ref{fig:scan}, and both $A_\lambda$ and $A_\kappa$ are
connected to the Higgs spectrum, as shown, \eg, by Eq.~\eqref{eq:mh1} and
Eq.~\eqref{eq:ma1}. Although deriving the exact relation between the $A_\lambda$
and $A_\kappa$ parameters that are defined in our setup at the grand unification
scale is a complex task due to renormalization group running, an
approximatively linear relation can be derived and is indeed observed in the
figure.

As a result of the above constraints, the typical spectrum exhibited by the
scenarios selected during our scan of the NMSSM parameter space features a
lightest neutralino state $\tilde\chi_1^0$ that is mainly singlino-like, whilst
the next two neutralinos $\tilde\chi_2^0$ and $\tilde\chi_3^0$ are close in mass
and higgsino-like. Once produced, these two heavier neutralinos can decay,
sometimes with a large branching ratio, into a final state system comprised of a
singlino $\tilde\chi_1^0$ and a pseudoscalar Higgs boson $A_1$. In the lower
right panel of Figure~\ref{fig:scan}, we present the dependence of these
branching ratios on the mass difference between the higgsino states (with masses
being approximatively taken as $\mu_{\rm eff}$) and the singlino.
This shows that an important $A_1$ production rate can be achieved in the region
where the spectrum is compressed, \ie, when the mass difference between the
higgsinos and the singlino is small. As soon as
$\mu_{\rm eff}-m_{\tilde\chi_1^0} \gtrsim 90$~GeV, the $\tilde{\chi}^0_i \to Z
\tilde{\chi}^0_1$ channel opens and quickly dominates. We have included in our
results the $\tilde\chi_3^0 \to A_1 \tilde\chi_2^0$
decay contributions. Although the phase space for such a decay process is very
limited as both higgsino states are close in mass (so that the $A_1$ decay
products are soft and very hard to detect), the final-state $\tilde\chi_2^0$
higgsino will further decay into an $A_1$ particle that will be energetic enough
to leave observable tracks in a detector.

\begin{figure}
  \centering
  \includegraphics[width=0.48\textwidth]{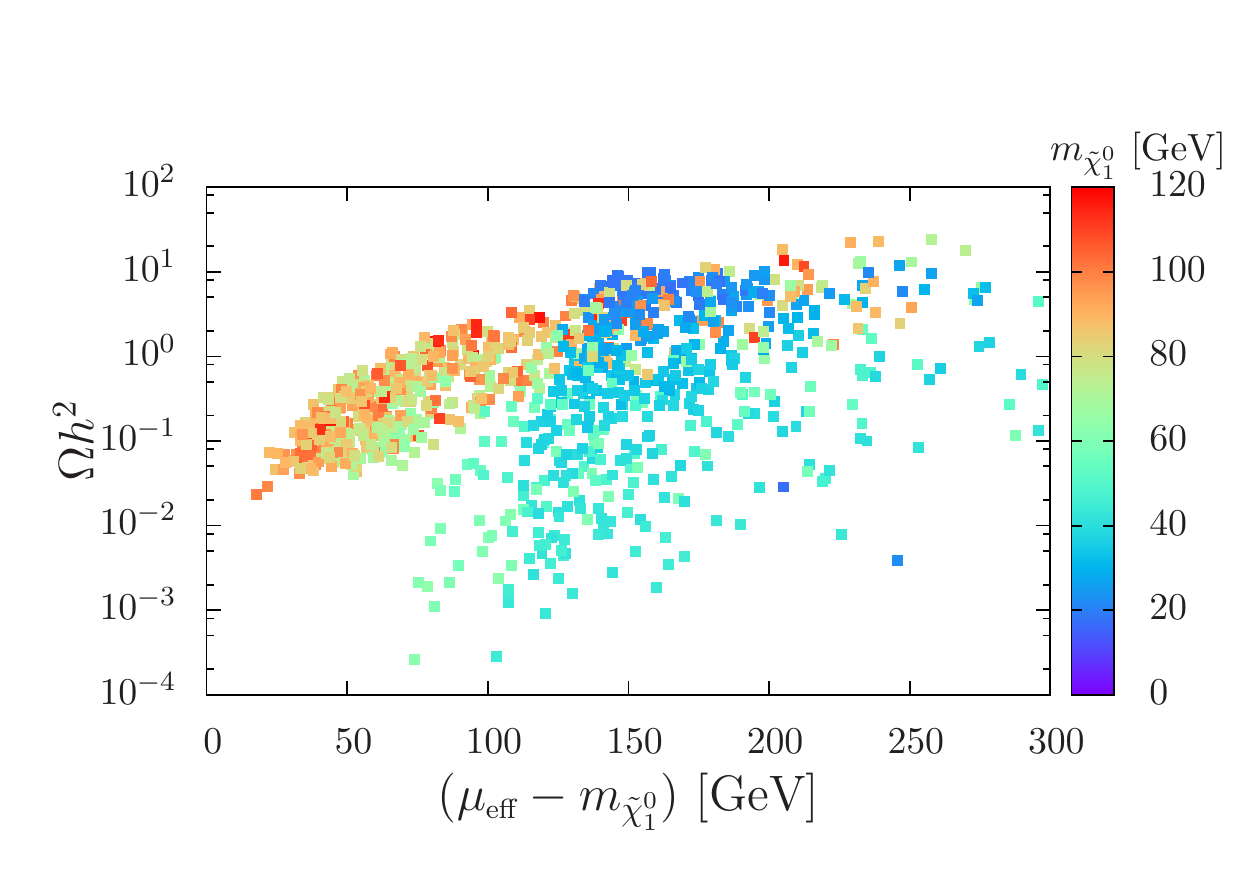}
  \includegraphics[width=0.48\textwidth]{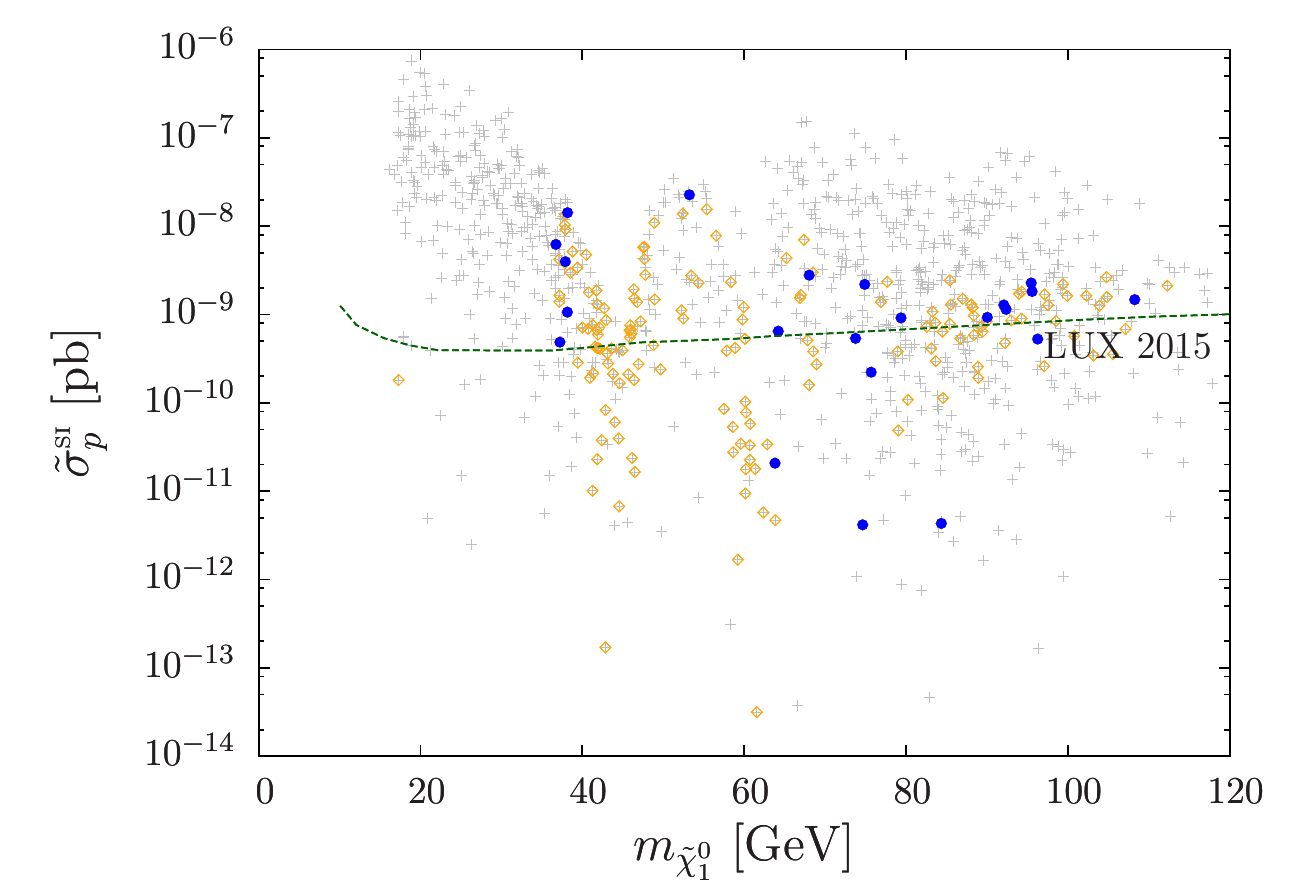}
  \caption{\label{fig:dm} Dark matter constraints that have been imposed during
    our NMSSM parameter space scan. We show, in the left panel, the dark matter
    relic density $\Omega h^2$ as a function of the mass splitting between the
    lightest (singlino-dominated) neutralino $\tilde\chi_1^0$ and the higgsinos,
    whose mass is of about  $\mu_{\rm eff}$. We additionally provide information    on the lightest neutralino mass (color code). On the right panel, we present
    the spin-independent dark-matter--proton scattering cross section scaled
    by $\Omega h^2 /0.119$ as a function of the lightest neutralino mass. We
    focus on scenarios for which $2 m_\tau < m_{A_1} < 2 m_b$ (gray) that
    additionally feature $\Omega h^2 < 0.131$ (orange) or
    $0.107 < \Omega h^2 < 0.131$ (blue). We indicate the LUX bounds from 2015 by
    a dark green line.}
\end{figure}

Additionally to the constraints that we have imposed so far, we moreover
restrict the properties of the lightest neutralino $\tilde\chi_1^0$ so that it
could be a good dark matter candidate. To this aim, we first impose that the
associated relic density abundance $\Omega h^2$ is compatible with the latest
Planck results~\cite{Ade:2015xua},
\be
  0.107 < \Omega h^2 < 0.131\ ,
\label{eq:Planck}\ee
where the allowed range for $\Omega h^2$ includes theoretical uncertainties on
the predictions. Since the $\tilde\chi_1^0$ particle is of a singlino-dominated
nature, a small $\mu - m_{\tilde{\chi}^0_1}$ mass splitting, or equivalently a
large higgsino-singlino mixing, is required for efficient enough dark matter
(co)annihilation processes. This is illustrated on the left panel of
Figure~\ref{fig:dm} in which
we study the relations between the lightest neutralino relic density as computed
with the {\sc MicrOmegas} package~\cite{Belanger:2005kh,Belanger:2013oya} and
the $\mu - m_{\tilde{\chi}^0_1}$ splitting. We observe that the scenarios
selected in our scan are spread among three regions of the parameter
space. In the heavy dark matter region for which $m_{\tilde{\chi}^0_1} \gtrsim
m_{H_2}/2$, dark matter annihilation proceeds via an $s$-channel $Z$-boson
exchange diagram that highly depends on the higgsino fraction of the lightest
neutralino $\tilde\chi_1^0$. As a consequence, a relic density in agreement with
Planck data implies that the mass splitting $\mu - m_{\tilde{\chi}^0_1}$ is at
most of about 70~GeV. In contrast, in the two other regions, the so-called Higgs
funnel region for which $m_{\tilde{\chi}^0_1} \sim m_{H_2}/2$ and $Z$-boson
funnel region for which $m_{\tilde{\chi}^0_1} \sim m_{Z}/2$, larger splittings
are allowed as the dark matter annihilation cross section is enhanced by the
presence of resonant diagrams that allow to recover a relic density in agreement
with Eq.~\eqref{eq:Planck}.

We next focus on the spin-independent dark-matter--proton scattering cross
section,
\be
  \tilde{\sigma}^{\text{SI}}_p=\sigma^{\text{SI}}_p \frac{\Omega h^2}{0.119}\ ,
\ee
and compare the NMSSM predictions obtained with {\sc MicrOmegas} to LUX
data~\cite{Akerib:2015rjg}. The results are
presented in the right panel of Figure~\ref{fig:dm}, in which we demonstrate
that many models in all three regions can survive all considered dark matter
constraints. The heavy dark matter region is additionally expected to be easier
to probe within future dark matter direct detection experiments that will
further constrain the scattering cross section $\tilde{\sigma}^{\text{SI}}_p$.

\subsection{Identification of benchmark scenarios for Run--II LHC studies}
\label{sec:benchmark}
All the NMSSM scenarios compliant with the constraints imposed so far exhibit
common features. The sfermions are typically lying in the multi-TeV region are
are thus
out of the reach of the LHC from the direct search standpoint, at least with an
assumed luminosity of 50--100~fb$^{-1}$. In contrast, the singlino and higgsino
states are in general light (with masses below 400--500~GeV) and the gauginos
heavy (with masses greater than 1~TeV). As a results, the only superpartners
that could be produced with a sufficiently large rate, and thus observed,
consist of the second and third neutralinos (or equivalently, the higgsino-like
neutralinos) and the lightest chargino. As already discussed in the previous
subsection, the singlet-like Higgs boson $H_1$ has a mass smaller than about
100~GeV, while the second scalar state $H_2$ is identified with the
Standard-Model Higgs boson. The lightest pseudoscalar state $A_1$ is
singlet-like with a mass satisfying $2 m_\tau < m_{A_1} < 2 m_b$ and all other
Higgs particles are beyond 1~TeV.

\begin{table}
\footnotesize
\center
  \setlength{\tabcolsep}{2.5mm}
  \renewcommand{\arraystretch}{1.4}
  \begin{tabular}{c|c|c|c|c||c|c|c|c}
   $m_0$ & $m_{1/2}$ & $A_0$ & $A_\lambda$ & $A_\kappa$&
   $\lambda$ & $\kappa$ & $\tan \beta$ & $\mu$\\ \hline
   1215.3~GeV & 1872.8~GeV & -4112.1~GeV & 301.1~GeV & 204.8~GeV&
   0.317 & 0.122 & 12.2 & 121.3~GeV
  \end{tabular}\\[.3cm]
  \begin{tabular}{c|c|c||c||c|c||c|c|c}
  $m_{\tilde{\chi}^0_1}$ & $m_{\tilde{\chi}^0_2}$ & $m_{\tilde{\chi}^0_3}$ &
    $m_{\tilde{\chi}^\pm_1}$ & $m_{A_1}$ & $m_{A_2}$ & $m_{H_1}$ & $m_{H_2}$ &
    $m_{H_3}$ \\ \hline
  75.7~GeV & -135.3~GeV & 149.2~GeV & 124.2~GeV & 5.5~GeV & 1538~GeV & 93.8~GeV
    & 125.9~GeV & 1538~GeV
  \end{tabular}\\[.3cm]
  \begin{tabular}{c||c|c||c||c}
    Br$({\tilde{\chi}^0_2\to A_1 \tilde{\chi}^0_1})$ &
    Br$({\tilde{\chi}^0_3\to A_1 \tilde{\chi}^0_1})$ &
    Br$({\tilde{\chi}^0_3\to A_1 \tilde{\chi}^0_2})$ &
    Br$({A_1 \to \tau \tau})$ & Br$({H_2 \to A_1 A_1})$\\
      98.9\% & 12.9\% & 87.1\% & 93.6\% & 4.2\%
  \end{tabular}\\[.3cm]
  \begin{tabular}{c|c||c|c}
    $\mu_{gg\to H,\gamma \gamma}$ & $\mu_{{\rm VBF}, VV^*}$ &
     $\Omega h^2$ &  $\sigma^{\text{SI}}_p$ \\  \hline
      1.06 & 1.02 &  0.107 & $2.46 \times 10^{-10}$~pb
  \end{tabular}
  \caption{\label{tab:bp} Representative NMSSM benchmark scenario satisfying all
    Higgs and dark matter constraints imposed in our scan. The scenario is
    defined in  the upper part of the table, in which we recall that the
    parameters of the left part of the table are given at the grand unification
    scale while those of its right part are defined at the electroweak scale.
    The light state masses are given in the middle panel of the table, while
    relevant branching ratios and obsevable results are given in its lower
    parts.
  }
\end{table}

In order to study the discovery prospects for these scenarios at the LHC, we
focus on a specific benchmark point that is defined in terms of the nine
parameters given in the upper panel of Table~\ref{tab:bp}. The light part of the
resulting mass spectrum is presented in the second panel of the table, while
relevant branching ratios and Higgs and dark matter observables are shown in
its two lower panels. In the context of such a benchmark scenario, the new
physics processes yielding the largest production cross section consist of
the production of a pair of (neutral or charged) higgsinos at the LHC. By a
virtue of their dominant decay modes, current LHC searches are not sensitive to
their associated signature. Neutralino production leads to the further
production
of pseudoscalar Higgs bosons that next decay into pairs of boosted taus, while
the lightest chargino decays into an off-shell $W$-boson and the singlino so
that no LHC bound is expected from the charged higgsino side~\cite{Aad:2014vma}.
Following the procedure developed in Ref.~\cite{Cheng:2013fma,Guo:2013asa}, we
recast all relevant Run--I LHC analyses. We find the CMS search for same-sign
dilepton~\cite{Chatrchyan:2013fea} is the most sensitive to NMSSM benchmark
points such as the one depicted above. However, the relevant production rates
are almost two orders of magnitude lower than the signal cross section that is
excluded at the 95\% confidence level.

In addition, the $H_2 \to A_1 A_1$ branching ratio reaches about 4\%, a result
compatible with Higgs current data. Although this decay mode has been actively
searched for~\cite{Khachatryan:2015nba,Aad:2015oqa,Khachatryan:2015wka} as it
consists of an irrefutable proof of the realization of an NMSSM
scenario~\cite{Ellwanger:2003jt,Ellwanger:2013ova}, we focus instead, in the
next section, on a novel discovery mode for a light NMSSM pseudoscalar based on
boosted ditau tagging.

\section{LHC sensitivity to light NMSSM pseudoscalar Higgs bosons decaying to a
boosted ditau jet}
\label{sec:simu}

\subsection{Event simulation for the NMSSM and boosted ditau tagging}
To determine the LHC sensitivity to the class of NMSSM models introduced in the
previous section, we analyze Monte Carlo simulations of proton-proton collisions
at a center-of-mass energy of 13~TeV as they could occur at the LHC collider at
CERN. Hard-scattering signal and background simulations rely on the
{\sc MadGraph5\_aMC@NLO}
program~\cite{Alwall:2014hca} that contains an NMSSM implementation. The latter,
that has not been described in any earlier publication, is generated by the
{\sc FeynRules} package~\cite{Alloul:2013bka} and its superspace
module~\cite{Duhr:2011se} that automatically produce a UFO
library~\cite{Degrande:2011ua} that can be employed by {\sc MadGraph5\_aMC@NLO}
for event generation, following the strategy of Ref.~\cite{Christensen:2009jx}.
The {\sc FeynRules} model includes a more general version of the superpotential
of Eq.~\eqref{eq:wnmssm} and the soft supersymmetry-breaking Lagrangian of
Eq.~\eqref{eq:lsoft} where the $\mathbb{Z}_3$ symmetry that we have imposed
in this work is not included and where intergenerational sfermion mixings are
allowed. Although such mixings are not handled by {\sc NmssmTools}, they are
compatible with the Supersymmetry Les Houches Accord
conventions~\cite{Skands:2003cj,Allanach:2008qq} in which their implementation
consists of
optional requirements. The translation of the output spectrum files produced by
{\sc NmssmTools} to files compliant with the NMSSM UFO is nevertheless
immediate as this only necessitates to increase the size of $2\times 2$ mixing
matrices to $6\times 6$ matrices with zero entries whenever two different
sfermion generations are concerned. The validation of the NMSSM implementation
of {\sc FeynRules} has been extensively performed during the 2009 Les Houches
workshop on TeV collider physics~\cite{Butterworth:2010ym}, and thousands of
supersymmetric processes have been considered to this aim.

The decays of the produced hard particles and the matching of the
parton-level hard events to a parton shower and hadronization infrastructure
have been performed in the context of the
{\sc Pythia}~6 package~\cite{Sjostrand:2006za},
that is interfaced to {\sc Tauola}~\cite{Jadach:1993hs,Davidson:2010rw}
for handling tau lepton decays. In this framework, any tau lepton polarization
effect is
neglected. Finally, we have simulated the response of the ATLAS detector by
means of the {\sc Delphes 3} program~\cite{deFavereau:2013fsa}, that internally
reconstructs objects on the basis of the anti-$k_T$ jet
algorithm~\cite{Cacciari:2008gp} as implemented in the {\sc FastJet}
software~\cite{Cacciari:2011ma}.

\begin{figure}
 \includegraphics[width=0.48\textwidth]{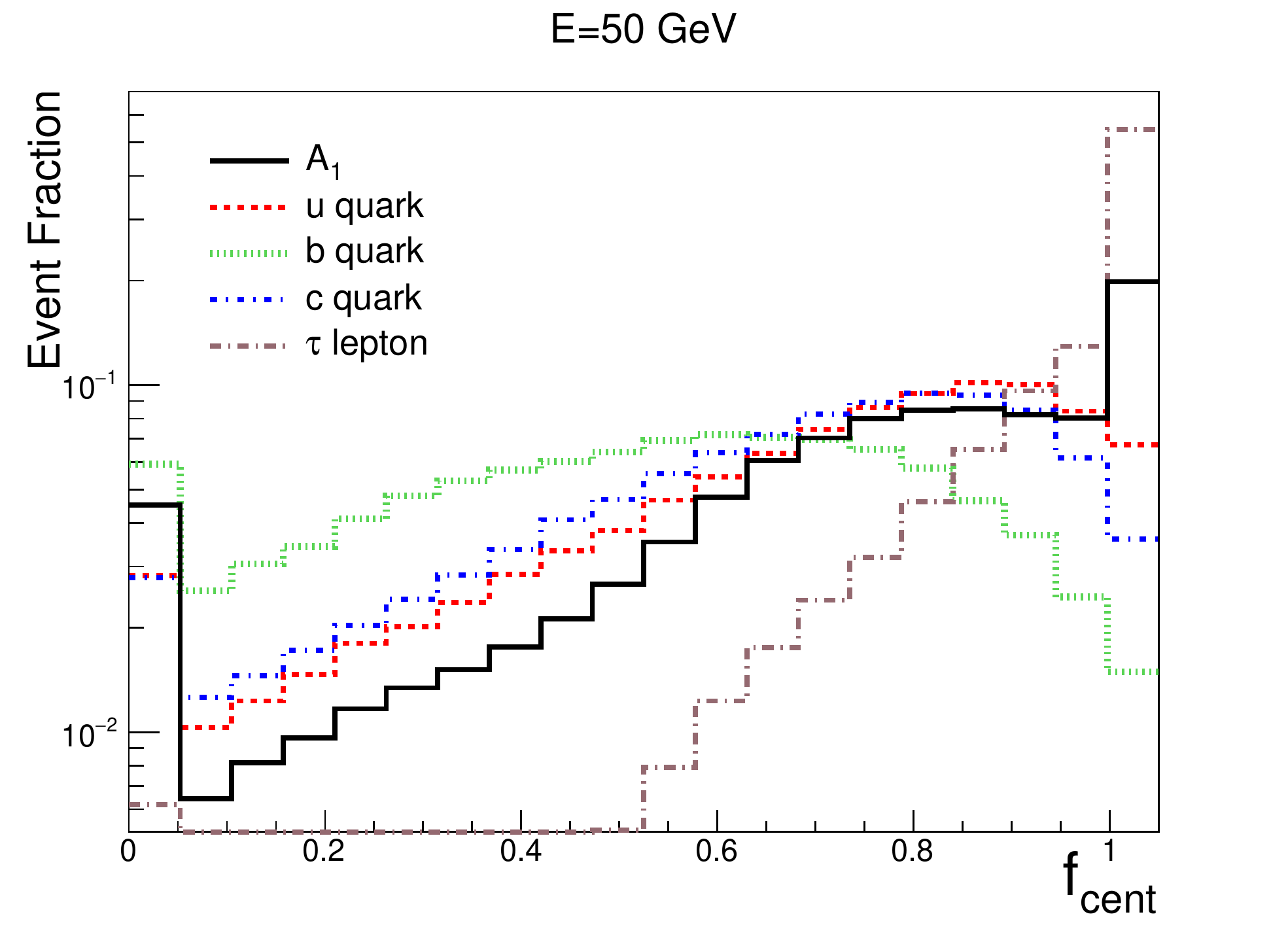}
 \includegraphics[width=0.48\textwidth]{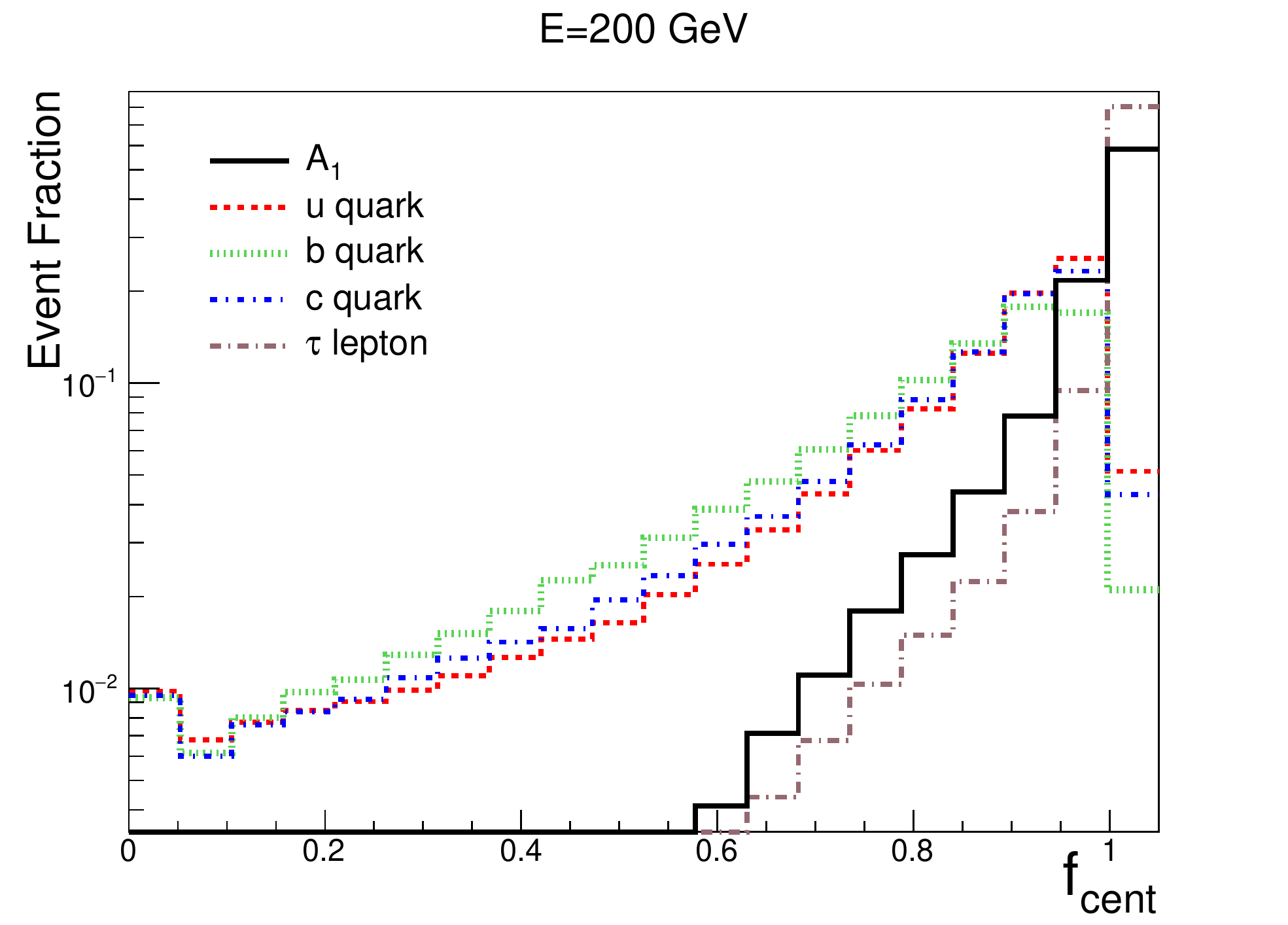} \\
 \includegraphics[width=0.48\textwidth]{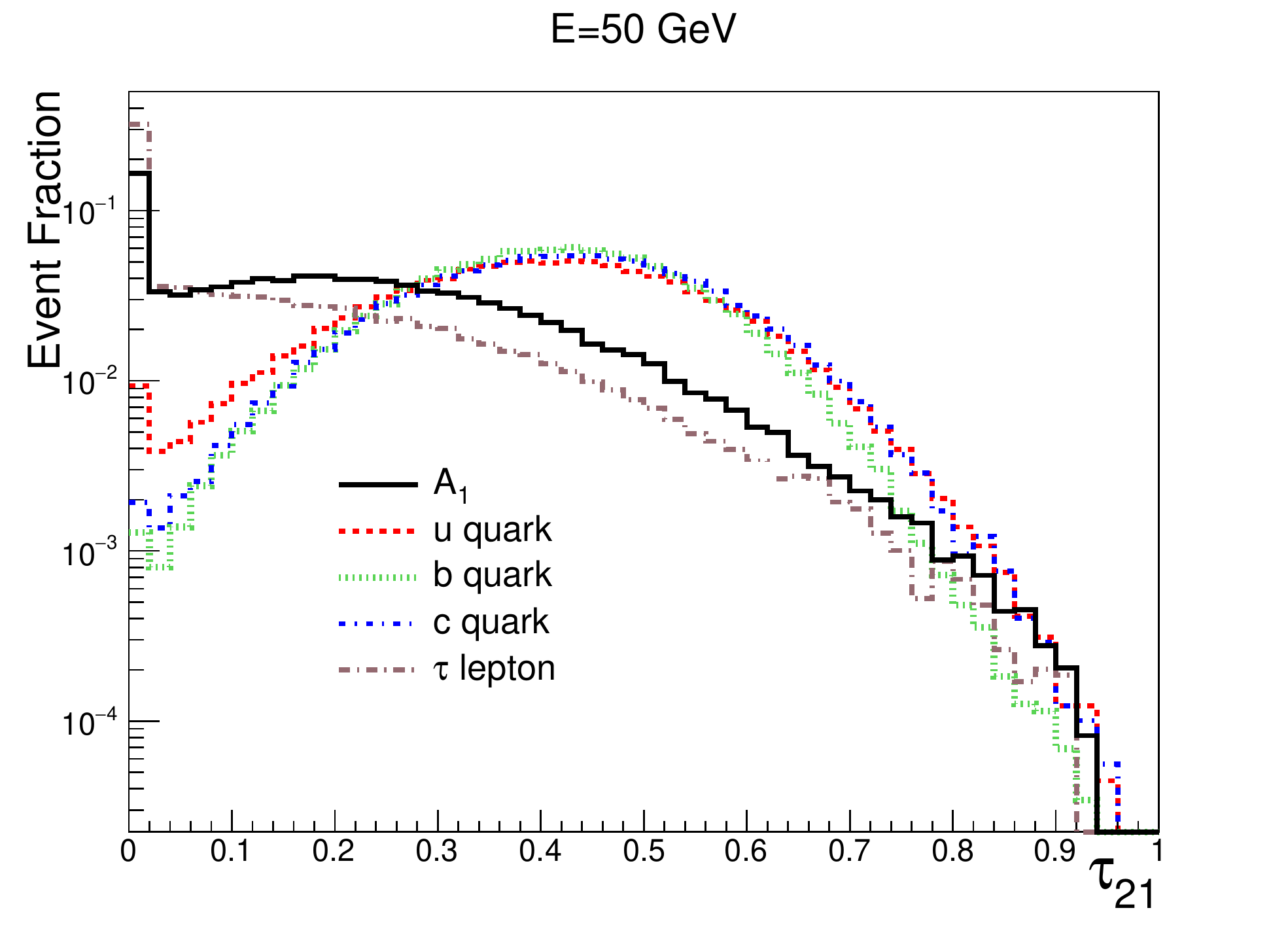}
 \includegraphics[width=0.48\textwidth]{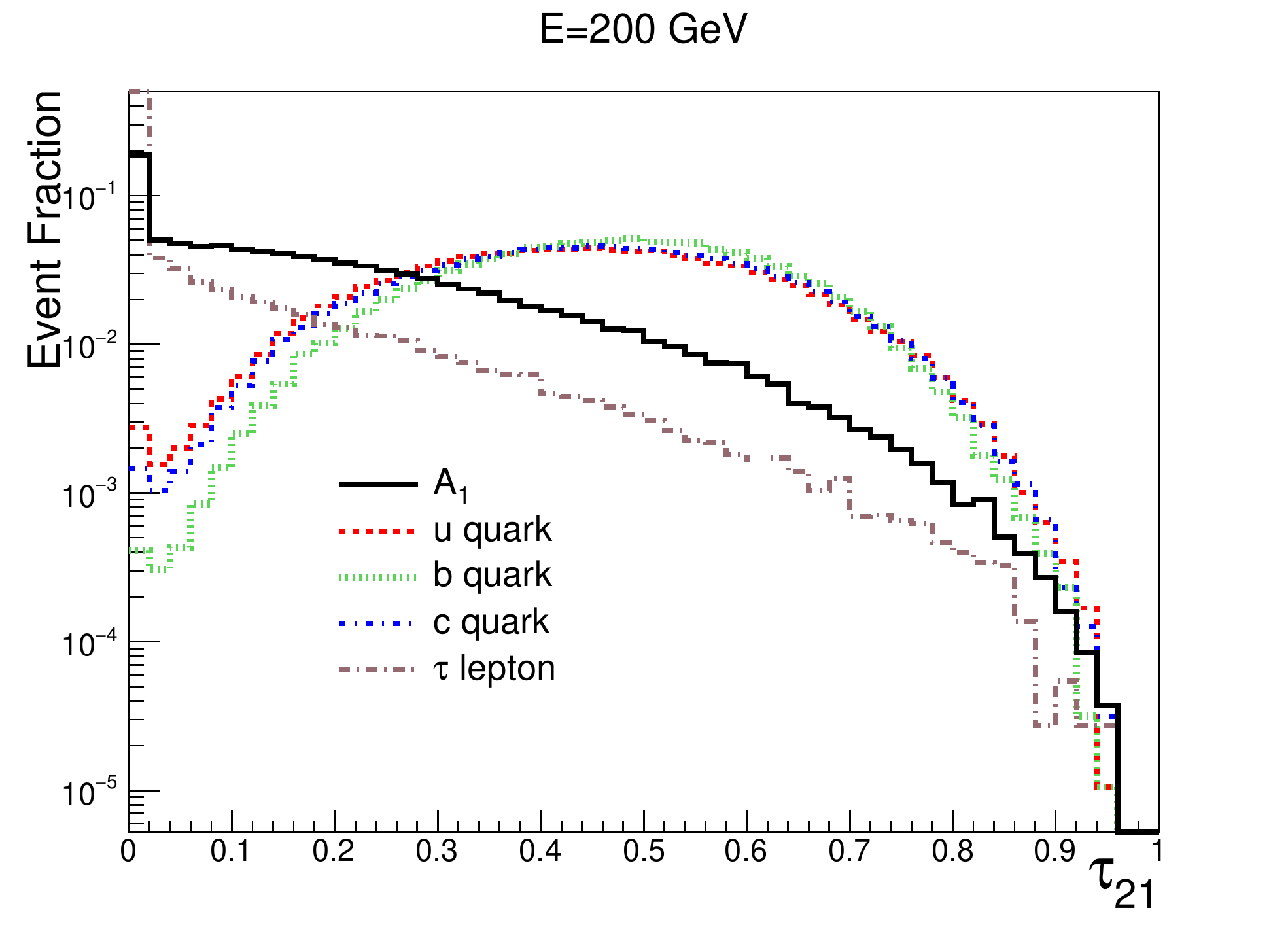}
 \caption{\label{fig:a1e} $f_{\text{cent}}$ (upper figures) and $\tau_{21}$
  (lower figures) distributions for jets originating from the fragmentation of
  up quarks (red dashed curves), charmed quarks (blue
  dash-dotted curves), bottom quarks (green dotted curves), a single tau lepton
  (brown dash-dotted curves) and a pseudoscalar NMSSM Higgs boson (black
  plain curves). Two jet energies of 50~GeV (left) and 200~GeV (right) are
  considered.}
\end{figure}

Since tau leptons dominantly decay hadronically with a corresponding branching
ratio of about 65\%, the development of efficient related tagging techniques is
a very important task in particular with respect to Higgs precision
measurements~\cite{Aad:2012mea} and new physics searches~\cite{Li:2015sza}. Jets
originating from pure QCD subprocesses and from the hadronic decay of a tau
lepton are mainly distinguished from each other by the number of charged tracks
inside the jet and the jet energy density profile. The properties specific to a
tau jet however turn out
to be preserved in the case of a boosted object comprised of two hadronically
decaying tau leptons that could arise from the decay of a heavier state.
Motivated by such considerations, we have designed an analysis strategy allowing
us to detect the signature of boosted pseudoscalar NMSSM Higgs bosons produced
from the decay of heavier higgsino states at the LHC. This relies on the tagging
of boosted ditau objects via a multivariate method~\cite{Aad:2014rga} that makes
use of the number of tracks inside the ditau jet, the ratio
$f_{\text{cent}}=E^{(0.1)}/E^{(0.2)}$ where $E^{(0.1)}$ ($E^{(0.2)}$) is the
total calorimetric energy deposit in a cone of radius $R=0.1$ (0.2) centered on
the jet direction, the transverse momentum of the hardest track inside a cone of
radius $R=0.2$ centered on the jet direction computed relatively to the jet
$p_T$, the $p_T$-weighted sum of the angular distances of all tracks inside the
jet, the maximum angular distance in the transverse plane between any track
lying inside a cone of radius $R=0.2$ centered on the jet direction and the jet
direction, the track-based jet invariant mass and the ratio of the jet
transverse momentum to the jet invariant mass.

This strategy can furthermore be improved to
gain sensitivity to leptonically decaying taus (within a boosted ditau
object) as well. The produced leptons are indeed unlikely to be isolated,
so that they could be captured by a selection involving the ratio of the
electromagnetic to hadronic calorimetric deposits. Additionally, we also
consider in our multivariate tagging technique the $N$-subjettiness variable
$\tau_{21} = \tau_2/\tau_1$ that allows one order to resolve the substructure of
the ditau jet~\cite{Kim:2010uj,Thaler:2010tr}, with $\tau_N$ being defined by
\be
  \tau_N =\frac{\sum_k \min \{ \Delta R_{1,k}, \Delta R_{2,k}, \dots, \Delta R_{N,k} \}}{\sum_k p_{T,k} R_0}.
\ee
In this expression, the summations have to be considered upon all jet
constituents, $R_0=0.4$ is the jet cone size parameter in the original jet clustering algorithm and $\Delta R_{I,k}$ denotes the distance in the
transverse plane between the subjet candidate $I$ and the jet constituent $k$.

\begin{figure}
  \includegraphics[width=0.48\textwidth]{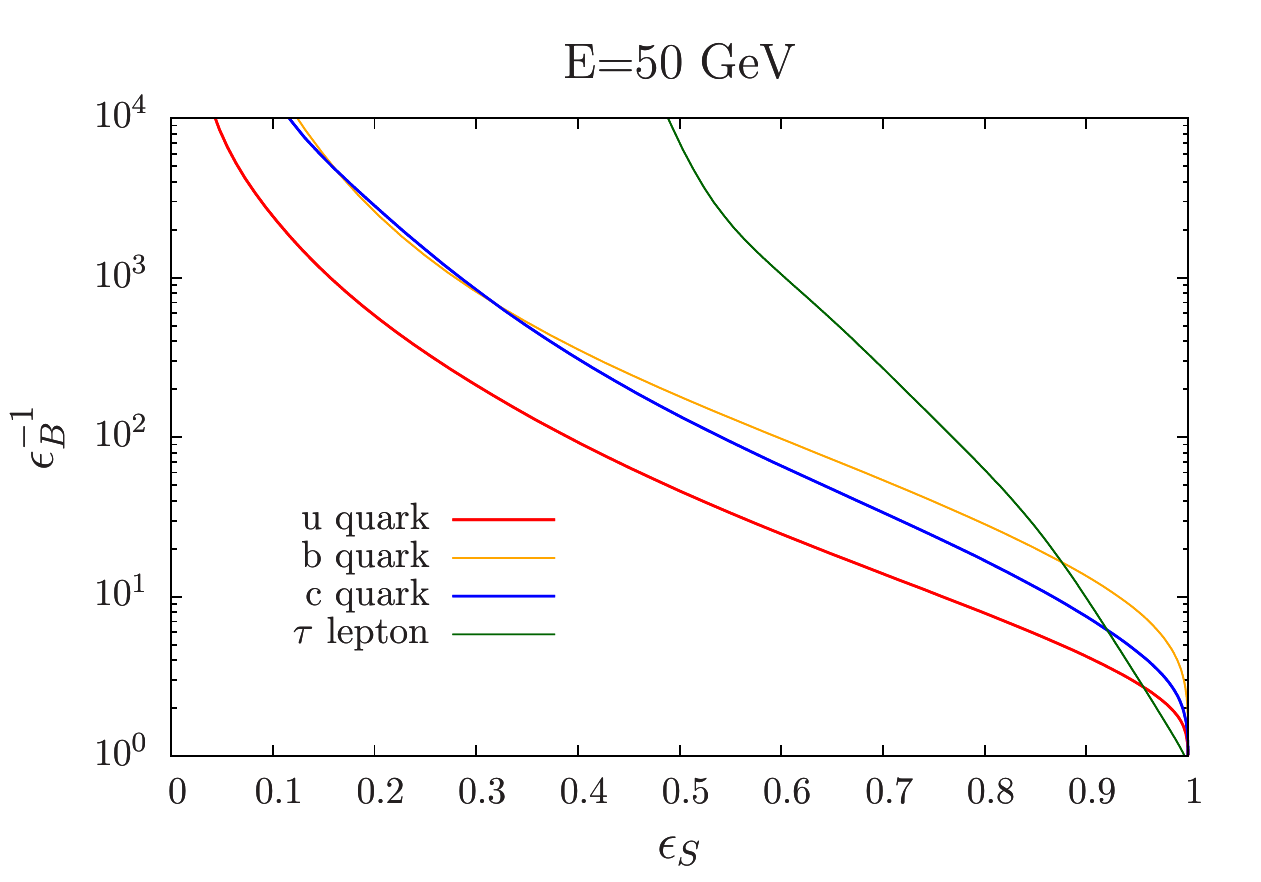}
  \includegraphics[width=0.48\textwidth]{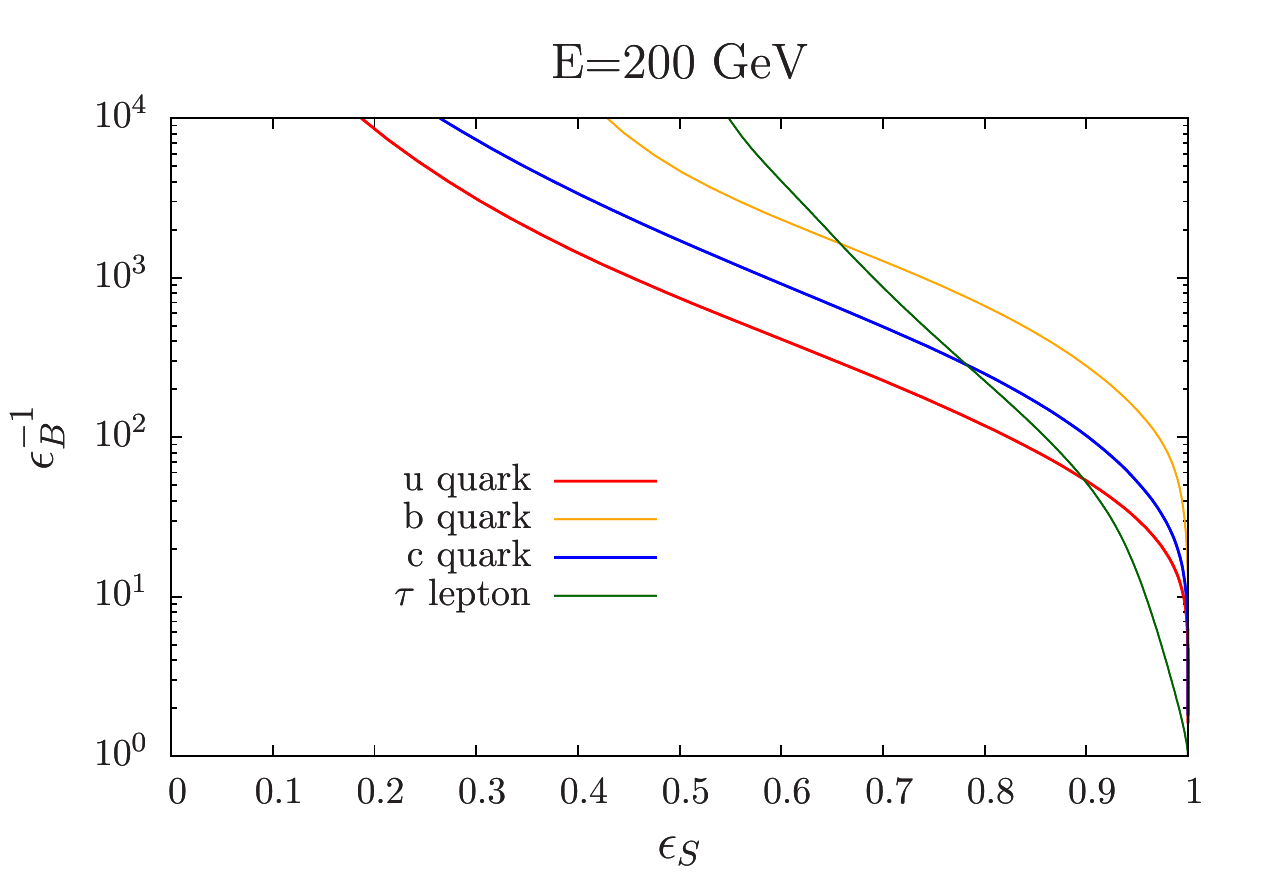}
  \caption{\label{fig:bdta1} Dependence of the quark and tau lepton rejection
    powers on the boosted ditau object tagging efficiency for jets of 50~GeV
    (left) and 200~GeV (right).}
\end{figure}

All the variables that we have introduced to tag a ditau boosted object strongly
depend on the object energy, as it is illustrated on Figure~\ref{fig:a1e} for
two representative jet energies of 50~GeV and 200~GeV and the $f_{\text{cent}}$
and $\tau_{21}$ variables. We compare the spectra that would be obtained when
jets solely originate from up quarks (red dashed curves), charmed quarks
(blue dash-dotted curves), bottom quarks (green dotted curves), a single tau
lepton (brown dash-dotted curves) and a pseudoscalar NMSSM Higgs boson
(black plain curves). The properties of ditau boosted objects are
different from the single tau jet and the purely QCD jet cases, so that there
exist handles for discriminating them. To this aim, we use a
boosted decision tree (BDT) technique that uses all the variables presented in
this section. The BDT is trained in the context of jets with specific energies
of 50~GeV and 200~GeV and the correlations between the obtained boosted
ditau object tagging efficiency and the QCD jet or single tau jet mistagging
rates are shown in Figure~\ref{fig:bdta1}. Jets issued from the fragmentation of
light quarks are always harder to distinguish from ditau boosted objects, as
their properties are similar to the ditau case (see Figure~\ref{fig:a1e}).
The corresponding
rejection power is particularly small when the jet energy is smaller. Taking
as a benchmark a tagging efficiency $\epsilon_S=50\%$, a rejection power
$\epsilon_B^{-1}$ of only 50 is found for 50~GeV jets, this number increasing to
500 for 200~GeV jets.

\subsection{LHC sensitivity to NMSSM light higgsinos decaying to
 boosted ditau objects}

In order to estimate the sensitivity of the LHC to the class of NMSSM scenarios
under consideration, we study the associated production of a
neutralino and a chargino, followed by a neutralino decay into a pseudoscalar
Higgs boson and a chargino decay into a far off-shell $W$-boson,
\be
  p p \to \tilde{\chi}^\pm_1 \tilde{\chi}^0_j \to (W^* \tilde{\chi}^0_1) 
    \ (A_1 \tilde{\chi}^0_1) \ .
\ee
This process features a large cross section, as this was already the case in the
MSSM~\cite{Debove:2008nr,Debove:2009ia,Debove:2011xj,Debove:2010kf,Fuks:2012qx,%
Fuks:2013vua}, and we further impose the off-shell $W$-boson to decay
leptonically so that fully hadronic backgrounds can be suppressed. In our
simulation, we use next-to-leading signal cross sections that are derived with
{\sc Prospino}~\cite{Beenakker:1999xh} and that mostly agree with the most
precise results involving soft-gluon resummation~\cite{Debove:2010kf,%
Fuks:2012qx,Fuks:2013vua}. The dominant associated sources of background
consist of events issued from the production of a (leptonically decaying)
$W$-boson in association with jets, of top-antitop systems and of diboson
systems that all give rise to
final states comprised of a single lepton, missing energy and hard jets. We
generate events exhibiting at least one jet at the
matrix element level and in which the hard-scattering lepton and jets have a
transverse momentum $p_T>10$~GeV and 20~GeV respectively, together with a
pseudorapidity satisfying
$|\eta|<2.5$. We normalize the top-antitop and diboson event samples to the
measured~\cite{xtt} and next-to-leading order~\cite{Campbell:2011bn} cross
section values respectively, and we make use of the leading-order $W$-boson plus
jet fiducial cross section as returned by {\sc MadGraph5\_aMC@NLO} to which a
next-to-leading order $K$-factor of 1.4 is included~\cite{Alwall:2014hca}.

\begin{figure}
  \includegraphics[width=0.48\textwidth]{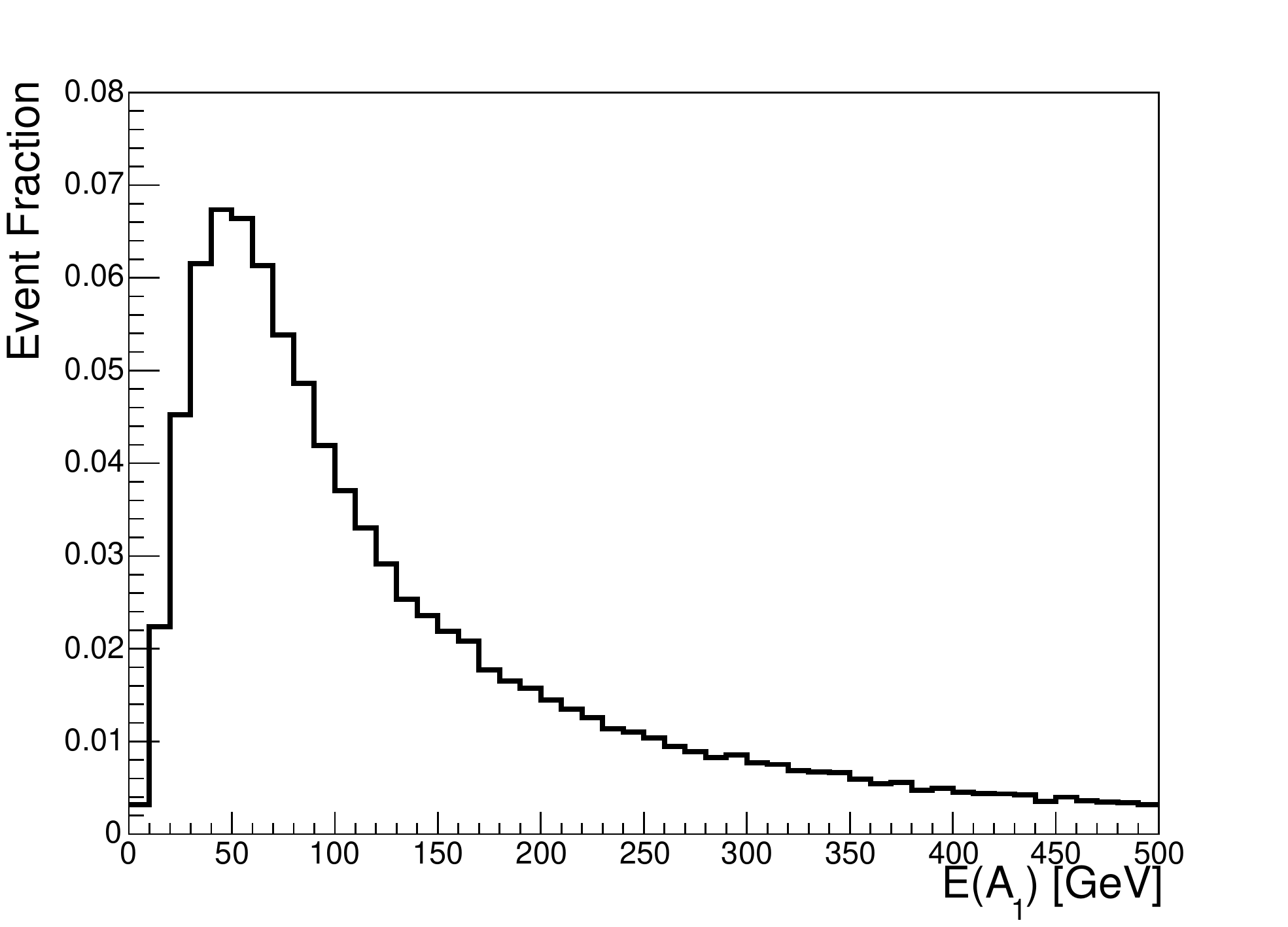}
  \includegraphics[width=0.48\textwidth]{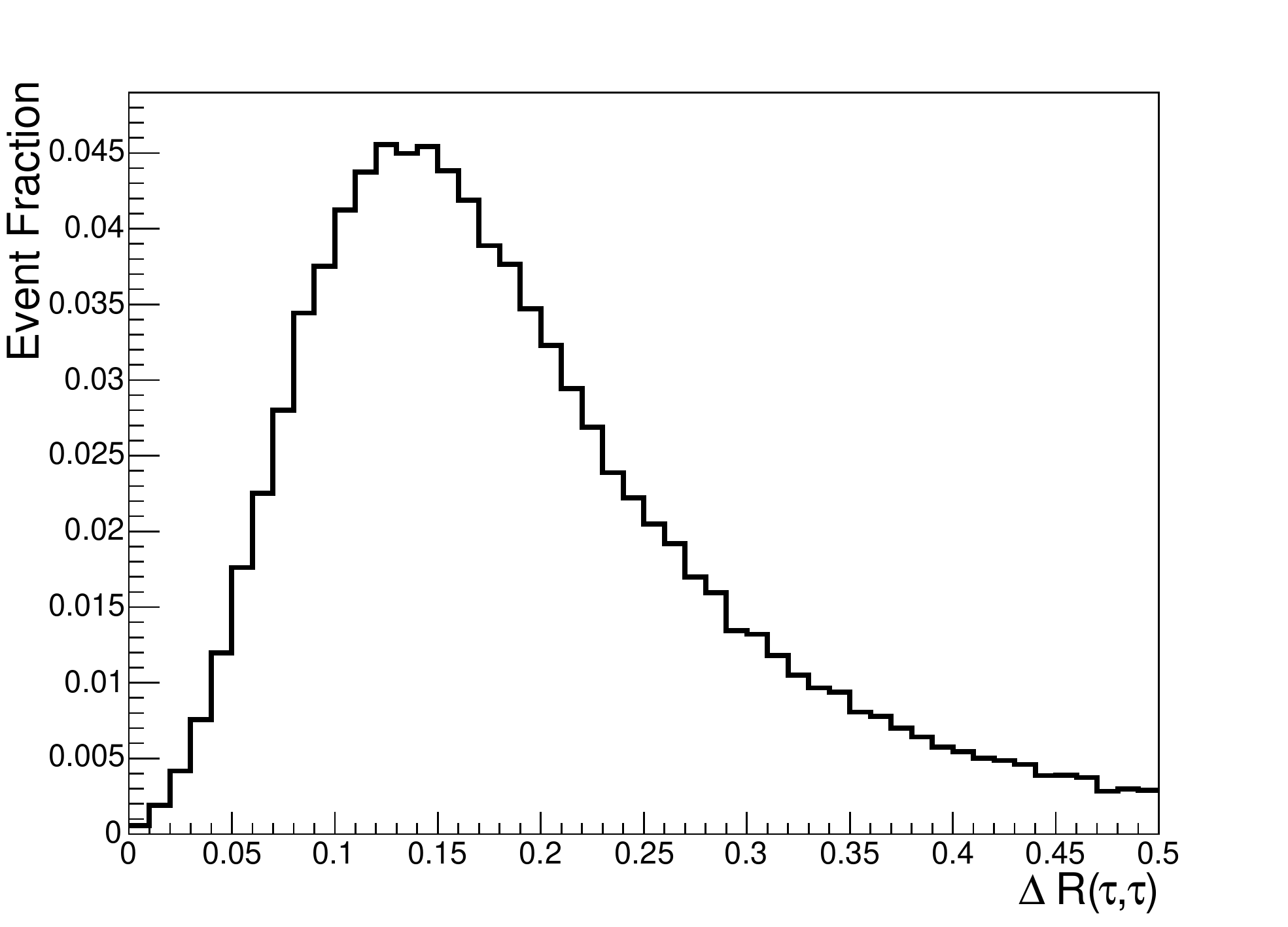}
  \caption{\label{fig:ea} Energy spectrum of the pseudoscalar Higgs bosons $A_1$
  issued from the decay of neutralino states (left) and the angular distance in
  the transverse plane between the two taus originating from the $A_1$ decays
  (right) in the context of signal events. Both distributions have been
  calculated for the NMSSM benchmark scenario introduced in
  Section~\ref{sec:benchmark} and are normalized to 1.}
\end{figure}

\begin{table}
\center
  \setlength{\tabcolsep}{2.5mm}
  \renewcommand{\arraystretch}{1.4}
  \begin{tabular}{c||c|ccc}
   & $\tilde{\chi}^\pm \tilde{\chi}^0$ signal  & $W$ plus jets &
    Top pair production & Diboson production \\ \hline\hline
  $\sigma^{13}$ & 3.38~pb & 8452~pb & 825~pb & 159.3~pb \\ \hline
  $\sigma^{\text{pre}}$ & 0.42~pb & 4313~pb & 62.9~pb & 29.2~pb \\
  \end{tabular}
  \caption{Signal and background cross section before ($\sigma^{13}$) and after
   ($\sigma^{\text{pre}}$) event preselection at 13 TeV LHC. \label{tab:pre}}
 \end{table}

Since signal events feature a final state comprised of a single lepton, a
boosted ditau object issued from the decay of a pseudoscalar $A_1$ particle and
missing energy, we preselect events by requiring that the final state contains
exactly one isolated lepton, at least one jet and we veto the presence of
$b$-tagged jets.
The resulting signal and background cross sections before ($\sigma^{13}$)
and after ($\sigma^{\text{pre}}$) the preselection at 13 TeV LHC are given in
Table~\ref{tab:pre}. At this stage of our analysis, background rates are of
about three orders of
magnitude larger than typical signal cross sections. In the NMSSM scenarios
under consideration, the spectrum generally features a mass splitting $\Delta M=
\mu_{\rm eff}-m_{\tilde\chi_1^0}$ between the lightest singlino-like neutralino
and the heavier neutral and charged higgsino states of at most 100~GeV (see the
lower right panel of Figure~\ref{fig:scan}), so that signal events generally
exhibit final state systems for which the lepton and jet transverse momenta and
the missing transverse energy are of about 50~GeV. It is consequently not
straightforward to design appropriate selections to enhance the signal over
background ratio by only means of the kinematical properties of the signal. The
mass splitting $\Delta M$ also determines the energy $E(A_1)$ that is typically
carried by the pseudoscalar Higgs bosons $A_1$ originating from the decays of
the higgsino-like neutralinos. Since $E(A_1)$ is in general much larger than
the $A_1$ mass, the pseudoscalar decay products (two tau leptons) turn out to
be highly collimated. This is illustrated, for the benchmark scenario introduced
in Section~\ref{sec:benchmark}, on Figure~\ref{fig:ea} where we present the
pseudoscalar Higgs boson $A_1$ energy spectrum (left panel) and the distribution
of the angular distance, in the transverse plane, between the two tau leptons
issued from the $A_1$ decays (right panel). %
\begin{figure}
 \includegraphics[width=0.48\textwidth]{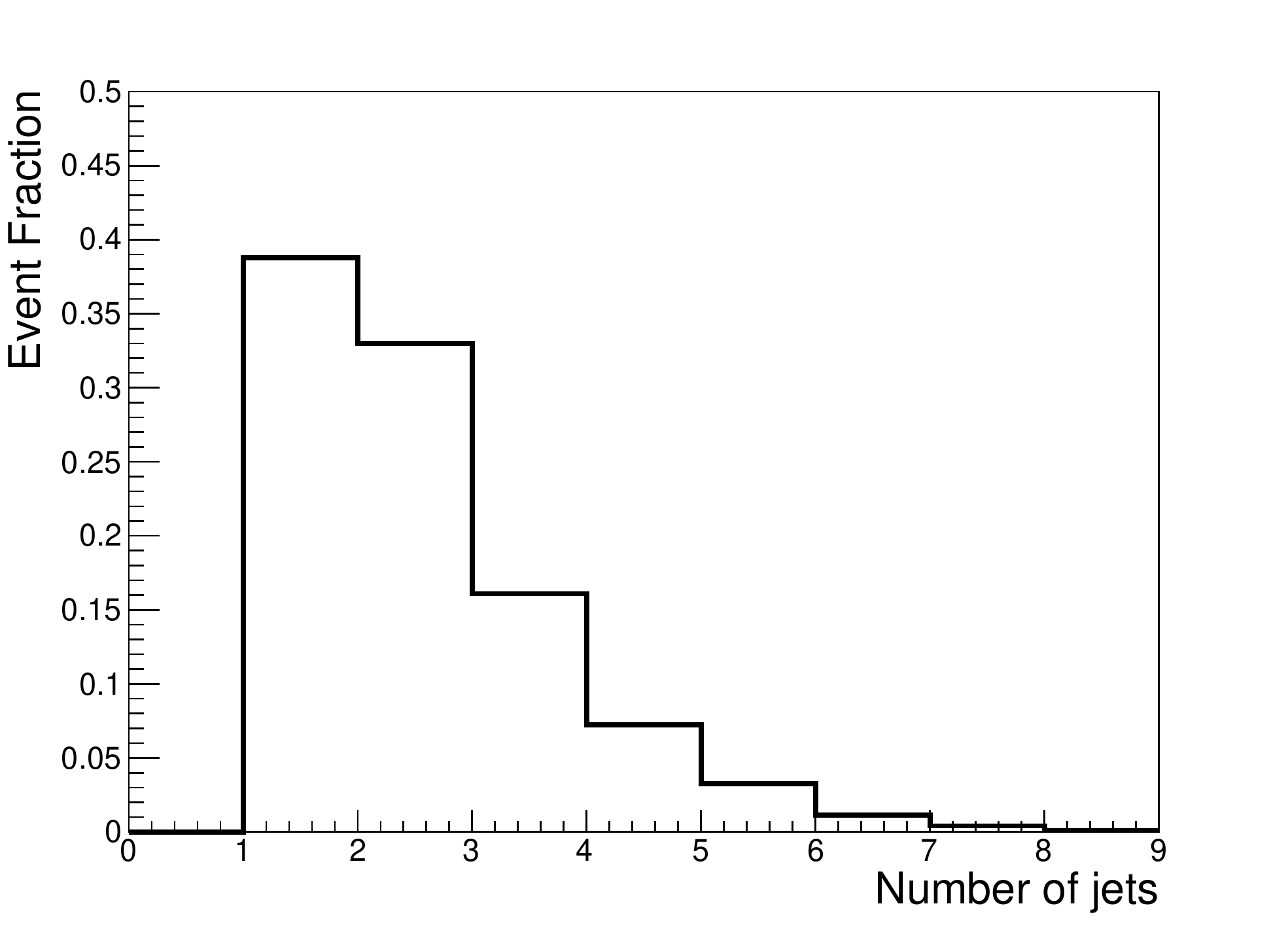}
 \includegraphics[width=0.48\textwidth]{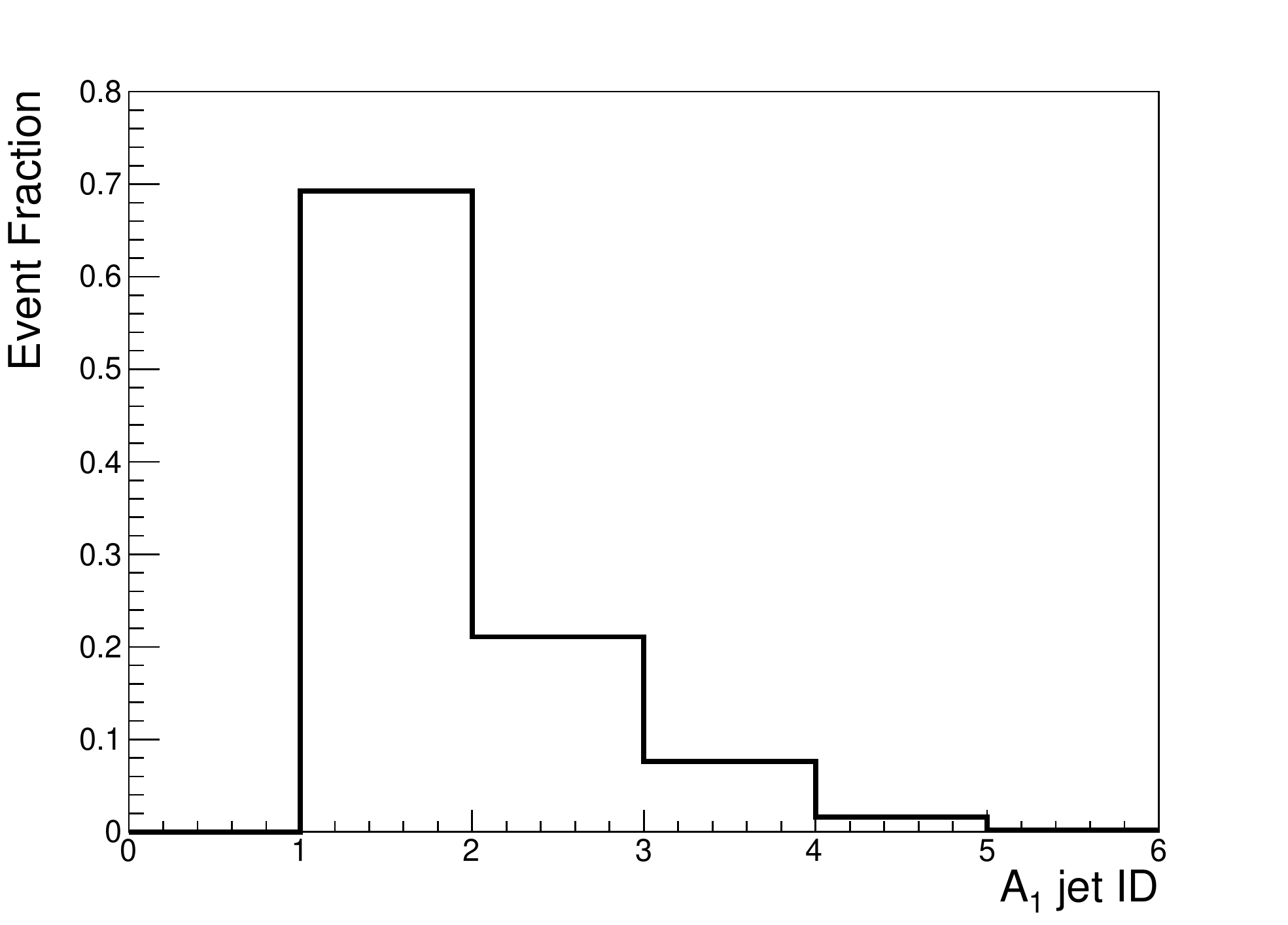}
 \caption{\label{fig:nj} The distribution in the number of jets (left) and the
   ranking of the boosted ditau jet associated with the $A_1$ particle (right)
   in the case of signal events and in the context of the benchmark scenario
   introduced in Section~\ref{sec:benchmark}. Both distributions are
   normalized to 1.}
\end{figure} %
In Figure~\ref{fig:nj}, we study the
details of the hadronic activity in the signal events. In the left panel of the
figure, we present the distribution of the number of jets $N_j$ characterizing
the signal events. Although only one jet (the boosted ditau object)
is expected from the partonic process, initial state radiation allows the $N_j$
spectrum to extent to larger values. The bulk of the events however features at
most two jets, while the leading jet is in general the boosted ditau object (in
70\% of the cases for the benchmark scenario of Section~\ref{sec:benchmark}).
This last property is depicted on the right panel of Figure~\ref{fig:nj}
where we present, on the basis of the Monte Carlo truth, the ranking of the
$A_1$ jet once the $p_T$ ordering of the jets is imposed.

As a consequence of these considerations, we further select signal events on the
basis of a multivariate analysis that uses as inputs the number of jets $N_j$,
the transverse momenta of the two leading jets $p_T^{j_1}$ and $p_T^{j_2}$, the
transverse
momentum of the lepton $p_T^\ell$, the invariant mass of the leading jet, the
amount of missing energy $\slashed{E}_T$, the angular distance in the azimuthal
direction with respect to the beam between the lepton and the missing momentum
$\Delta \phi_{\ell,\slashed{E}_T}$, and the angular distance in the transverse
plane between the lepton and the leading jet $\Delta R(\ell,j_1)$. In addition,
we also include in our analysis the reconstructed $W$-boson transverse mass
$m_T^W$ that would be obtained when considering that all the missing transverse
energy is connected to a $W$-boson decay,
\be
  M_T^W = \sqrt{2 p_T^\ell \slashed{E}_T \Big[1-\cos \Delta \phi_{\ell,\slashed{E}_T} \Big]} \ ,
\ee
and the final state stranverse mass $m_{T2}$~\cite{Lester:1999tx,Cheng:2008hk}.
Both these latter variables are expected to provide a handle to efficiently
suppress the dominant $W$ background. Moreover, assuming that the boosted
ditau object is identified with the leading jet, the next-to-leading jet is
expected to be softer in the signal case than in the background case,
the lepton and missing energy tend to be not correlated in the signal case as
there are several sources of missing energy, and both the $M_T^W$ and $M_{T2}$
variables are distributed towards smaller values for the signal as the
considered signature is free from any on-shell $W$ boson. The analysis can
finally be improved after imposing that the leading jet is a boosted ditau
object.

\begin{figure}
  \includegraphics[width=0.48\textwidth]{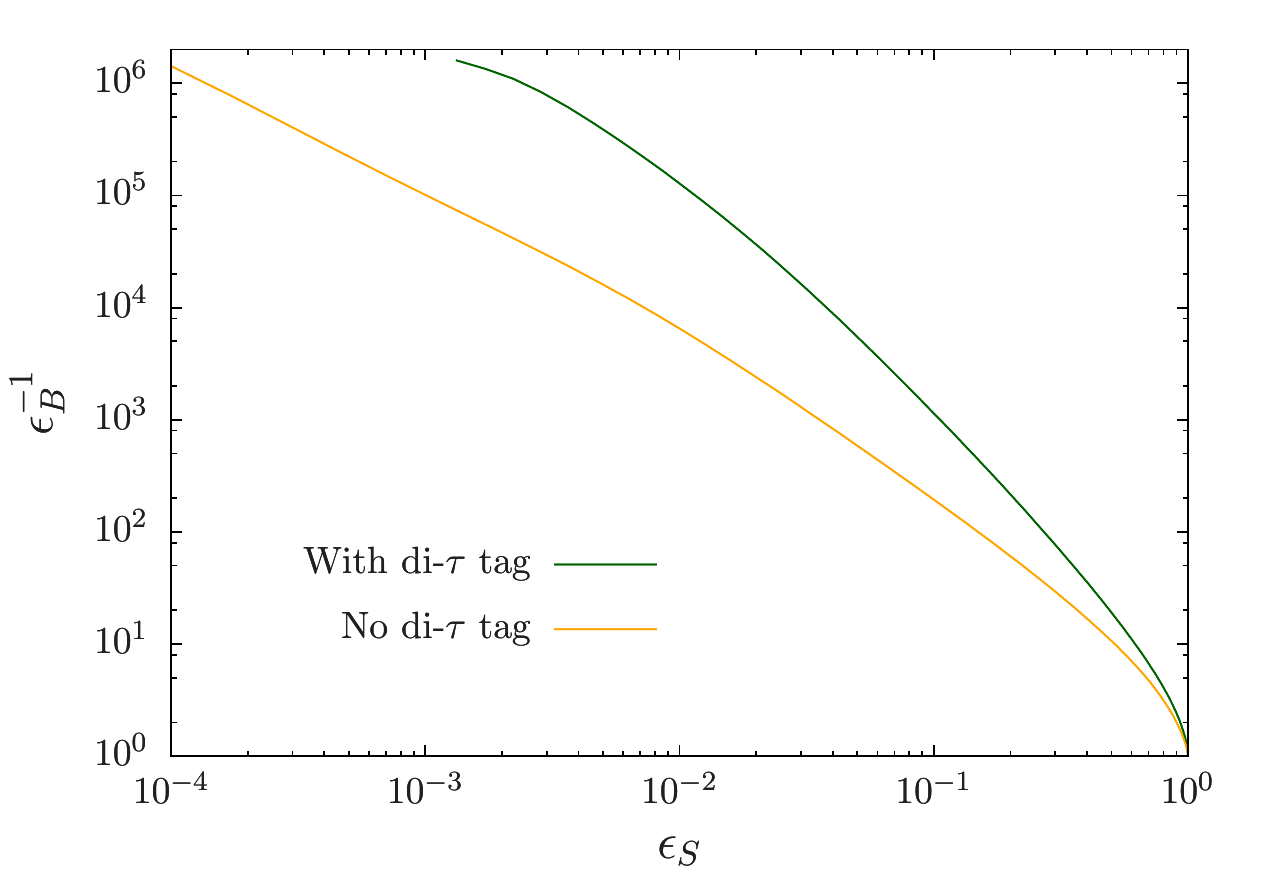}
  \includegraphics[width=0.48\textwidth]{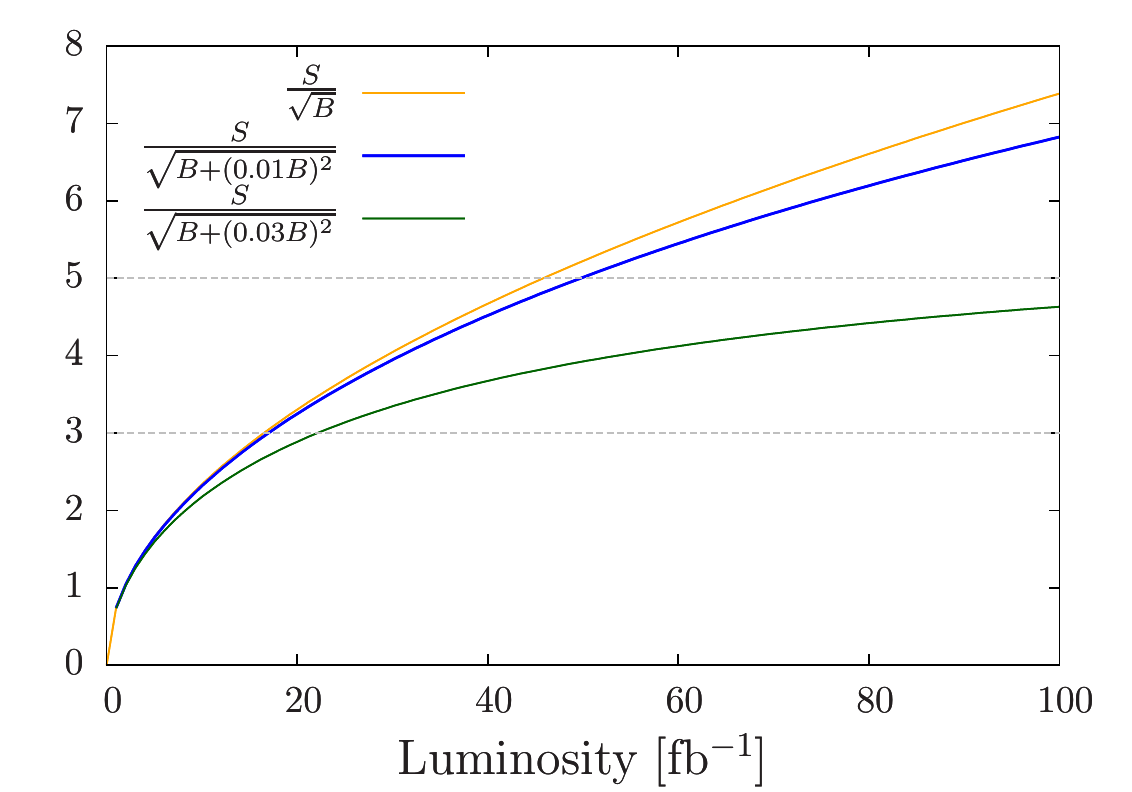}
  \caption{\label{fig:sig} Left: relations between the signal selection
   efficiency and background rejection rate obtained by means of our
   multivariate
   analysis technique. The effect of the tagging of the leading jet as a boosted
   ditau object is either included (dark green) or not (orange). Right: signal
   significance obtained with our analysis as a function of the luminosity and
   for different treatments of the systematic uncertainties on the background.}
\end{figure}

Applying a selection on the BDT response, we can derive the dependence of the
background rejection rate $\epsilon^{-1}_B$ on the signal selection efficiency
$\epsilon_S$, as shown in the left panel of Figure~\ref{fig:sig}. We
consider two different setups in which the boosted ditau object feature of the
signal is accounted for (dark green) or not (orange), and we observe
that an efficiency of about one percent can be obtained together with a background
rejection rate of $10^4$ ($\gtrsim 10^5$) when the ditau tagging is ignored
(included). Optimizing the selection on the BDT output, the resulting signal fiducial
cross section is of 3.1~fb, while the dominant $W$-boson plus jet and diboson
background component ones are of 16.4~fb and 0.9~fb respectively. Moreover, top-antitop
events turn to contribute negligibly, with a fiducial cross section smaller than
0.5~fb. We calculate the associated signal
significance $\sigma_{\rm signal}$ as a function of the luminosity for different
treatment of the systematic uncertainties on the background $\Delta B$,
\be
  \sigma_{\rm signal} = \frac{S}{\sqrt{B+(\Delta B)^2}} \ ,
\ee
and show the results on the right panel of the figure. In this expression, $S$ and $B$ denote the
number of selected signal and background events respectively. Assuming a
systematic uncertainty on the background at the percent level, a $3\sigma$ hint for the
class of NMSSM scenarios considered in this work could be observed at the
early stage of the LHC Run--II, while a $5\sigma$ discovery could be expected
with an integrated luminosity of at least 50~fb$^{-1}$. 

\section{Conclusion}
\label{sec:conl}
It has been recently shown that the discovery of a light pseudoscalar state
could be a direct evidence for the next-to-minimal realization of supersymmetry
in nature. As a result, many studies have been dedicated to the
investigation of the discovery potential associated with such particles. Most
existing works focus on heavier Higgs decay probes, although light pseudoscalar
states could also be copiously produced from neutralino decays.

In this work, we have explored the NMSSM parameter space and impose
Higgs and dark matter constraints on the construction of viable scenarios. We
have found that many of such scenarios include pseudoscalar Higgs bosons with a
mass comprised in the $[2 m_\tau, 2 m_b]$ window. In this case, the
golden discovery mode consists of the production of an associated
chargino-neutralino pair that further decays into a boosted ditau jet
(connected to a pseudoscalar decay), a single lepton and missing energy. We
study a typical reference scenario an investigate the sensitivity of the LHC
Run--II to the corresponding signal. By means of a multivariate analysis and a
boosted ditau object tagging method, we have found that the background could be
rejected at a very large level so that a $3\sigma$ hint for the signal is
expected within the first 13 TeV data, and that a $5\sigma$ discovery could be
envisaged for a luminosity of more than about 50~fb$^{-1}$.

\section*{Acknowledgements} 
This work has ben supported in parts by the {\it Th\'eorie LHC France}
initiative of the CNRS and a joint CNRS-CAS doctoral grant. The work of JL and AGW is supported by the Australian Research Council through the Centre of Excellence for Particle Physics at the Terascale CE110001004.

\bibliographystyle{JHEP}
\bibliography{lightA.bib}

\begin{thebibliography}{86}
\expandafter\ifx\csname natexlab\endcsname\relax\def\natexlab#1{#1}\fi
\expandafter\ifx\csname bibnamefont\endcsname\relax
  \def\bibnamefont#1{#1}\fi
\expandafter\ifx\csname bibfnamefont\endcsname\relax
  \def\bibfnamefont#1{#1}\fi
\expandafter\ifx\csname citenamefont\endcsname\relax
  \def\citenamefont#1{#1}\fi
\expandafter\ifx\csname url\endcsname\relax
  \def\url#1{\texttt{#1}}\fi
\expandafter\ifx\csname urlprefix\endcsname\relax\def\urlprefix{URL }\fi
\providecommand{\bibinfo}[2]{#2}
\providecommand{\eprint}[2][]{\url{#2}}

\bibitem[{\citenamefont{Nilles}(1984)}]{Nilles:1983ge}
\bibinfo{author}{\bibfnamefont{H.~P.} \bibnamefont{Nilles}},
  \bibinfo{journal}{Phys. Rept.} \textbf{\bibinfo{volume}{110}},
  \bibinfo{pages}{1} (\bibinfo{year}{1984}).

\bibitem[{\citenamefont{Haber and Kane}(1985)}]{Haber:1984rc}
\bibinfo{author}{\bibfnamefont{H.~E.} \bibnamefont{Haber}} \bibnamefont{and}
  \bibinfo{author}{\bibfnamefont{G.~L.} \bibnamefont{Kane}},
  \bibinfo{journal}{Phys. Rept.} \textbf{\bibinfo{volume}{117}},
  \bibinfo{pages}{75} (\bibinfo{year}{1985}).

\bibitem[{\citenamefont{Kim and Nilles}(1984)}]{Kim:1983dt}
\bibinfo{author}{\bibfnamefont{J.~E.} \bibnamefont{Kim}} \bibnamefont{and}
  \bibinfo{author}{\bibfnamefont{H.~P.} \bibnamefont{Nilles}},
  \bibinfo{journal}{Phys. Lett.} \textbf{\bibinfo{volume}{B138}},
  \bibinfo{pages}{150} (\bibinfo{year}{1984}).

\bibitem[{\citenamefont{Aad et~al.}(2012{\natexlab{a}})}]{Aad:2012tfa}
\bibinfo{author}{\bibfnamefont{G.}~\bibnamefont{Aad}} \bibnamefont{et~al.}
  (\bibinfo{collaboration}{ATLAS}), \bibinfo{journal}{Phys. Lett.}
  \textbf{\bibinfo{volume}{B716}}, \bibinfo{pages}{1}
  (\bibinfo{year}{2012}{\natexlab{a}}), \eprint{1207.7214}.

\bibitem[{\citenamefont{Chatrchyan et~al.}(2012)}]{Chatrchyan:2012xdj}
\bibinfo{author}{\bibfnamefont{S.}~\bibnamefont{Chatrchyan}}
  \bibnamefont{et~al.} (\bibinfo{collaboration}{CMS}), \bibinfo{journal}{Phys.
  Lett.} \textbf{\bibinfo{volume}{B716}}, \bibinfo{pages}{30}
  (\bibinfo{year}{2012}), \eprint{1207.7235}.

\bibitem[{\citenamefont{Ellwanger et~al.}(2010)\citenamefont{Ellwanger,
  Hugonie, and Teixeira}}]{Ellwanger:2009dp}
\bibinfo{author}{\bibfnamefont{U.}~\bibnamefont{Ellwanger}},
  \bibinfo{author}{\bibfnamefont{C.}~\bibnamefont{Hugonie}}, \bibnamefont{and}
  \bibinfo{author}{\bibfnamefont{A.~M.} \bibnamefont{Teixeira}},
  \bibinfo{journal}{Phys. Rept.} \textbf{\bibinfo{volume}{496}},
  \bibinfo{pages}{1} (\bibinfo{year}{2010}), \eprint{0910.1785}.

\bibitem[{\citenamefont{Ellwanger and Hugonie}(2016)}]{Ellwanger:2016qax}
\bibinfo{author}{\bibfnamefont{U.}~\bibnamefont{Ellwanger}} \bibnamefont{and}
  \bibinfo{author}{\bibfnamefont{C.}~\bibnamefont{Hugonie}}
  (\bibinfo{year}{2016}), \eprint{1602.03344}.

\bibitem[{\citenamefont{{The ATLAS
  collaboration}}(2015{\natexlab{a}})}]{ATLAS-CONF-2015-081}
\bibinfo{author}{\bibnamefont{{The ATLAS collaboration}}},
  \bibinfo{journal}{ATLAS-CONF-2015-081}  (\bibinfo{year}{2015}{\natexlab{a}}).

\bibitem[{\citenamefont{{The CMS collaboration}}(2015)}]{CMS:2015dxe}
\bibinfo{author}{\bibnamefont{{The CMS collaboration}}},
  \bibinfo{journal}{CMS-PAS-EXO-15-004}  (\bibinfo{year}{2015}).

\bibitem[{\citenamefont{Khachatryan
  et~al.}(2015{\natexlab{a}})}]{Khachatryan:2014jba}
\bibinfo{author}{\bibfnamefont{V.}~\bibnamefont{Khachatryan}}
  \bibnamefont{et~al.} (\bibinfo{collaboration}{CMS}), \bibinfo{journal}{Eur.
  Phys. J.} \textbf{\bibinfo{volume}{C75}}, \bibinfo{pages}{212}
  (\bibinfo{year}{2015}{\natexlab{a}}), \eprint{1412.8662}.

\bibitem[{\citenamefont{Aad et~al.}(2016)}]{Aad:2015gba}
\bibinfo{author}{\bibfnamefont{G.}~\bibnamefont{Aad}} \bibnamefont{et~al.}
  (\bibinfo{collaboration}{ATLAS}), \bibinfo{journal}{Eur. Phys. J.}
  \textbf{\bibinfo{volume}{C76}}, \bibinfo{pages}{6} (\bibinfo{year}{2016}),
  \eprint{1507.04548}.

\bibitem[{\citenamefont{Miller et~al.}(2004)\citenamefont{Miller, Nevzorov, and
  Zerwas}}]{Miller:2003ay}
\bibinfo{author}{\bibfnamefont{D.~J.} \bibnamefont{Miller}},
  \bibinfo{author}{\bibfnamefont{R.}~\bibnamefont{Nevzorov}}, \bibnamefont{and}
  \bibinfo{author}{\bibfnamefont{P.~M.} \bibnamefont{Zerwas}},
  \bibinfo{journal}{Nucl. Phys.} \textbf{\bibinfo{volume}{B681}},
  \bibinfo{pages}{3} (\bibinfo{year}{2004}), \eprint{hep-ph/0304049}.

\bibitem[{\citenamefont{Dermisek and Gunion}(2005)}]{Dermisek:2005ar}
\bibinfo{author}{\bibfnamefont{R.}~\bibnamefont{Dermisek}} \bibnamefont{and}
  \bibinfo{author}{\bibfnamefont{J.~F.} \bibnamefont{Gunion}},
  \bibinfo{journal}{Phys. Rev. Lett.} \textbf{\bibinfo{volume}{95}},
  \bibinfo{pages}{041801} (\bibinfo{year}{2005}), \eprint{hep-ph/0502105}.

\bibitem[{\citenamefont{Dermisek and Gunion}(2007)}]{Dermisek:2006wr}
\bibinfo{author}{\bibfnamefont{R.}~\bibnamefont{Dermisek}} \bibnamefont{and}
  \bibinfo{author}{\bibfnamefont{J.~F.} \bibnamefont{Gunion}},
  \bibinfo{journal}{Phys. Rev.} \textbf{\bibinfo{volume}{D75}},
  \bibinfo{pages}{075019} (\bibinfo{year}{2007}), \eprint{hep-ph/0611142}.

\bibitem[{\citenamefont{Ellwanger et~al.}(2005)\citenamefont{Ellwanger, Gunion,
  and Hugonie}}]{Ellwanger:2005uu}
\bibinfo{author}{\bibfnamefont{U.}~\bibnamefont{Ellwanger}},
  \bibinfo{author}{\bibfnamefont{J.~F.} \bibnamefont{Gunion}},
  \bibnamefont{and} \bibinfo{author}{\bibfnamefont{C.}~\bibnamefont{Hugonie}},
  \bibinfo{journal}{JHEP} \textbf{\bibinfo{volume}{07}}, \bibinfo{pages}{041}
  (\bibinfo{year}{2005}), \eprint{hep-ph/0503203}.

\bibitem[{\citenamefont{Djouadi et~al.}(2008)}]{Djouadi:2008uw}
\bibinfo{author}{\bibfnamefont{A.}~\bibnamefont{Djouadi}} \bibnamefont{et~al.},
  \bibinfo{journal}{JHEP} \textbf{\bibinfo{volume}{07}}, \bibinfo{pages}{002}
  (\bibinfo{year}{2008}), \eprint{0801.4321}.

\bibitem[{\citenamefont{Cao et~al.}(2013)\citenamefont{Cao, Ding, Han, Yang,
  and Zhu}}]{Cao:2013gba}
\bibinfo{author}{\bibfnamefont{J.}~\bibnamefont{Cao}},
  \bibinfo{author}{\bibfnamefont{F.}~\bibnamefont{Ding}},
  \bibinfo{author}{\bibfnamefont{C.}~\bibnamefont{Han}},
  \bibinfo{author}{\bibfnamefont{J.~M.} \bibnamefont{Yang}}, \bibnamefont{and}
  \bibinfo{author}{\bibfnamefont{J.}~\bibnamefont{Zhu}},
  \bibinfo{journal}{JHEP} \textbf{\bibinfo{volume}{11}}, \bibinfo{pages}{018}
  (\bibinfo{year}{2013}), \eprint{1309.4939}.

\bibitem[{\citenamefont{Bomark et~al.}(2015{\natexlab{a}})\citenamefont{Bomark,
  Moretti, Munir, and Roszkowski}}]{Bomark:2014gya}
\bibinfo{author}{\bibfnamefont{N.-E.} \bibnamefont{Bomark}},
  \bibinfo{author}{\bibfnamefont{S.}~\bibnamefont{Moretti}},
  \bibinfo{author}{\bibfnamefont{S.}~\bibnamefont{Munir}}, \bibnamefont{and}
  \bibinfo{author}{\bibfnamefont{L.}~\bibnamefont{Roszkowski}},
  \bibinfo{journal}{JHEP} \textbf{\bibinfo{volume}{02}}, \bibinfo{pages}{044}
  (\bibinfo{year}{2015}{\natexlab{a}}), \eprint{1409.8393}.

\bibitem[{\citenamefont{Bomark et~al.}(2015{\natexlab{b}})\citenamefont{Bomark,
  Moretti, and Roszkowski}}]{Bomark:2015fga}
\bibinfo{author}{\bibfnamefont{N.-E.} \bibnamefont{Bomark}},
  \bibinfo{author}{\bibfnamefont{S.}~\bibnamefont{Moretti}}, \bibnamefont{and}
  \bibinfo{author}{\bibfnamefont{L.}~\bibnamefont{Roszkowski}}
  (\bibinfo{year}{2015}{\natexlab{b}}), \eprint{1503.04228}.

\bibitem[{\citenamefont{Potter}(2015)}]{Potter:2015wsa}
\bibinfo{author}{\bibfnamefont{C.~T.} \bibnamefont{Potter}}
  (\bibinfo{year}{2015}), \eprint{1505.05554}.

\bibitem[{\citenamefont{Almarashi and Moretti}(2011)}]{Almarashi:2011te}
\bibinfo{author}{\bibfnamefont{M.}~\bibnamefont{Almarashi}} \bibnamefont{and}
  \bibinfo{author}{\bibfnamefont{S.}~\bibnamefont{Moretti}},
  \bibinfo{journal}{Phys. Rev.} \textbf{\bibinfo{volume}{D84}},
  \bibinfo{pages}{035009} (\bibinfo{year}{2011}), \eprint{1106.1599}.

\bibitem[{\citenamefont{Forshaw et~al.}(2008)\citenamefont{Forshaw, Gunion,
  Hodgkinson, Papaefstathiou, and Pilkington}}]{Forshaw:2007ra}
\bibinfo{author}{\bibfnamefont{J.~R.} \bibnamefont{Forshaw}},
  \bibinfo{author}{\bibfnamefont{J.~F.} \bibnamefont{Gunion}},
  \bibinfo{author}{\bibfnamefont{L.}~\bibnamefont{Hodgkinson}},
  \bibinfo{author}{\bibfnamefont{A.}~\bibnamefont{Papaefstathiou}},
  \bibnamefont{and} \bibinfo{author}{\bibfnamefont{A.~D.}
  \bibnamefont{Pilkington}}, \bibinfo{journal}{JHEP}
  \textbf{\bibinfo{volume}{04}}, \bibinfo{pages}{090} (\bibinfo{year}{2008}),
  \eprint{0712.3510}.

\bibitem[{\citenamefont{Belyaev et~al.}(2008)\citenamefont{Belyaev, Hesselbach,
  Lehti, Moretti, Nikitenko, and Shepherd-Themistocleous}}]{Belyaev:2008gj}
\bibinfo{author}{\bibfnamefont{A.}~\bibnamefont{Belyaev}},
  \bibinfo{author}{\bibfnamefont{S.}~\bibnamefont{Hesselbach}},
  \bibinfo{author}{\bibfnamefont{S.}~\bibnamefont{Lehti}},
  \bibinfo{author}{\bibfnamefont{S.}~\bibnamefont{Moretti}},
  \bibinfo{author}{\bibfnamefont{A.}~\bibnamefont{Nikitenko}},
  \bibnamefont{and} \bibinfo{author}{\bibfnamefont{C.~H.}
  \bibnamefont{Shepherd-Themistocleous}} (\bibinfo{year}{2008}),
  \eprint{0805.3505}.

\bibitem[{\citenamefont{Cerdeno et~al.}(2013)\citenamefont{Cerdeno, Ghosh, and
  Park}}]{Cerdeno:2013cz}
\bibinfo{author}{\bibfnamefont{D.~G.} \bibnamefont{Cerdeno}},
  \bibinfo{author}{\bibfnamefont{P.}~\bibnamefont{Ghosh}}, \bibnamefont{and}
  \bibinfo{author}{\bibfnamefont{C.~B.} \bibnamefont{Park}},
  \bibinfo{journal}{JHEP} \textbf{\bibinfo{volume}{06}}, \bibinfo{pages}{031}
  (\bibinfo{year}{2013}), \eprint{1301.1325}.

\bibitem[{\citenamefont{Belyaev et~al.}(2010)\citenamefont{Belyaev, Pivarski,
  Safonov, Senkin, and Tatarinov}}]{Belyaev:2010ka}
\bibinfo{author}{\bibfnamefont{A.}~\bibnamefont{Belyaev}},
  \bibinfo{author}{\bibfnamefont{J.}~\bibnamefont{Pivarski}},
  \bibinfo{author}{\bibfnamefont{A.}~\bibnamefont{Safonov}},
  \bibinfo{author}{\bibfnamefont{S.}~\bibnamefont{Senkin}}, \bibnamefont{and}
  \bibinfo{author}{\bibfnamefont{A.}~\bibnamefont{Tatarinov}},
  \bibinfo{journal}{Phys. Rev.} \textbf{\bibinfo{volume}{D81}},
  \bibinfo{pages}{075021} (\bibinfo{year}{2010}), \eprint{1002.1956}.

\bibitem[{\citenamefont{Lisanti and Wacker}(2009)}]{Lisanti:2009uy}
\bibinfo{author}{\bibfnamefont{M.}~\bibnamefont{Lisanti}} \bibnamefont{and}
  \bibinfo{author}{\bibfnamefont{J.~G.} \bibnamefont{Wacker}},
  \bibinfo{journal}{Phys. Rev.} \textbf{\bibinfo{volume}{D79}},
  \bibinfo{pages}{115006} (\bibinfo{year}{2009}), \eprint{0903.1377}.

\bibitem[{\citenamefont{Ellwanger et~al.}(2003)\citenamefont{Ellwanger, Gunion,
  Hugonie, and Moretti}}]{Ellwanger:2003jt}
\bibinfo{author}{\bibfnamefont{U.}~\bibnamefont{Ellwanger}},
  \bibinfo{author}{\bibfnamefont{J.~F.} \bibnamefont{Gunion}},
  \bibinfo{author}{\bibfnamefont{C.}~\bibnamefont{Hugonie}}, \bibnamefont{and}
  \bibinfo{author}{\bibfnamefont{S.}~\bibnamefont{Moretti}}
  (\bibinfo{year}{2003}), \eprint{hep-ph/0305109}.

\bibitem[{\citenamefont{Ellwanger}(2013)}]{Ellwanger:2013ova}
\bibinfo{author}{\bibfnamefont{U.}~\bibnamefont{Ellwanger}},
  \bibinfo{journal}{JHEP} \textbf{\bibinfo{volume}{08}}, \bibinfo{pages}{077}
  (\bibinfo{year}{2013}), \eprint{1306.5541}.

\bibitem[{\citenamefont{Khachatryan
  et~al.}(2015{\natexlab{b}})}]{Khachatryan:2015nba}
\bibinfo{author}{\bibfnamefont{V.}~\bibnamefont{Khachatryan}}
  \bibnamefont{et~al.} (\bibinfo{collaboration}{CMS})
  (\bibinfo{year}{2015}{\natexlab{b}}), \eprint{1510.06534}.

\bibitem[{\citenamefont{Aad et~al.}(2015{\natexlab{a}})}]{Aad:2015oqa}
\bibinfo{author}{\bibfnamefont{G.}~\bibnamefont{Aad}} \bibnamefont{et~al.}
  (\bibinfo{collaboration}{ATLAS}), \bibinfo{journal}{Phys. Rev.}
  \textbf{\bibinfo{volume}{D92}}, \bibinfo{pages}{052002}
  (\bibinfo{year}{2015}{\natexlab{a}}), \eprint{1505.01609}.

\bibitem[{\citenamefont{Khachatryan et~al.}(2016)}]{Khachatryan:2015wka}
\bibinfo{author}{\bibfnamefont{V.}~\bibnamefont{Khachatryan}}
  \bibnamefont{et~al.} (\bibinfo{collaboration}{CMS}), \bibinfo{journal}{Phys.
  Lett.} \textbf{\bibinfo{volume}{B752}}, \bibinfo{pages}{146}
  (\bibinfo{year}{2016}), \eprint{1506.00424}.

\bibitem[{\citenamefont{Cheung and Hou}(2009)}]{Cheung:2008rh}
\bibinfo{author}{\bibfnamefont{K.}~\bibnamefont{Cheung}} \bibnamefont{and}
  \bibinfo{author}{\bibfnamefont{T.-J.} \bibnamefont{Hou}},
  \bibinfo{journal}{Phys. Lett.} \textbf{\bibinfo{volume}{B674}},
  \bibinfo{pages}{54} (\bibinfo{year}{2009}), \eprint{0809.1122}.

\bibitem[{\citenamefont{Cerdeño et~al.}(2014)\citenamefont{Cerdeño, Ghosh,
  Park, and Peiró}}]{Cerdeno:2013qta}
\bibinfo{author}{\bibfnamefont{D.~G.} \bibnamefont{Cerdeño}},
  \bibinfo{author}{\bibfnamefont{P.}~\bibnamefont{Ghosh}},
  \bibinfo{author}{\bibfnamefont{C.~B.} \bibnamefont{Park}}, \bibnamefont{and}
  \bibinfo{author}{\bibfnamefont{M.}~\bibnamefont{Peiró}},
  \bibinfo{journal}{JHEP} \textbf{\bibinfo{volume}{02}}, \bibinfo{pages}{048}
  (\bibinfo{year}{2014}), \eprint{1307.7601}.

\bibitem[{\citenamefont{Han et~al.}(2015)\citenamefont{Han, Kim, Munir, and
  Park}}]{Han:2015zba}
\bibinfo{author}{\bibfnamefont{C.}~\bibnamefont{Han}},
  \bibinfo{author}{\bibfnamefont{D.}~\bibnamefont{Kim}},
  \bibinfo{author}{\bibfnamefont{S.}~\bibnamefont{Munir}}, \bibnamefont{and}
  \bibinfo{author}{\bibfnamefont{M.}~\bibnamefont{Park}},
  \bibinfo{journal}{JHEP} \textbf{\bibinfo{volume}{07}}, \bibinfo{pages}{002}
  (\bibinfo{year}{2015}), \eprint{1504.05085}.

\bibitem[{\citenamefont{Englert et~al.}(2011)\citenamefont{Englert, Roy, and
  Spannowsky}}]{Englert:2011iz}
\bibinfo{author}{\bibfnamefont{C.}~\bibnamefont{Englert}},
  \bibinfo{author}{\bibfnamefont{T.~S.} \bibnamefont{Roy}}, \bibnamefont{and}
  \bibinfo{author}{\bibfnamefont{M.}~\bibnamefont{Spannowsky}},
  \bibinfo{journal}{Phys. Rev.} \textbf{\bibinfo{volume}{D84}},
  \bibinfo{pages}{075026} (\bibinfo{year}{2011}), \eprint{1106.4545}.

\bibitem[{\citenamefont{Papaefstathiou
  et~al.}(2014)\citenamefont{Papaefstathiou, Sakurai, and
  Takeuchi}}]{Papaefstathiou:2014oja}
\bibinfo{author}{\bibfnamefont{A.}~\bibnamefont{Papaefstathiou}},
  \bibinfo{author}{\bibfnamefont{K.}~\bibnamefont{Sakurai}}, \bibnamefont{and}
  \bibinfo{author}{\bibfnamefont{M.}~\bibnamefont{Takeuchi}},
  \bibinfo{journal}{JHEP} \textbf{\bibinfo{volume}{08}}, \bibinfo{pages}{176}
  (\bibinfo{year}{2014}), \eprint{1404.1077}.

\bibitem[{\citenamefont{Katz et~al.}(2011)\citenamefont{Katz, Son, and
  Tweedie}}]{Katz:2010iq}
\bibinfo{author}{\bibfnamefont{A.}~\bibnamefont{Katz}},
  \bibinfo{author}{\bibfnamefont{M.}~\bibnamefont{Son}}, \bibnamefont{and}
  \bibinfo{author}{\bibfnamefont{B.}~\bibnamefont{Tweedie}},
  \bibinfo{journal}{Phys. Rev.} \textbf{\bibinfo{volume}{D83}},
  \bibinfo{pages}{114033} (\bibinfo{year}{2011}), \eprint{1011.4523}.

\bibitem[{\citenamefont{Olive et~al.}(2014)}]{Agashe:2014kda}
\bibinfo{author}{\bibfnamefont{K.~A.} \bibnamefont{Olive}} \bibnamefont{et~al.}
  (\bibinfo{collaboration}{Particle Data Group}), \bibinfo{journal}{Chin.
  Phys.} \textbf{\bibinfo{volume}{C38}}, \bibinfo{pages}{090001}
  (\bibinfo{year}{2014}).

\bibitem[{\citenamefont{Ellwanger and Hugonie}(2007)}]{Ellwanger:2006rn}
\bibinfo{author}{\bibfnamefont{U.}~\bibnamefont{Ellwanger}} \bibnamefont{and}
  \bibinfo{author}{\bibfnamefont{C.}~\bibnamefont{Hugonie}},
  \bibinfo{journal}{Comput. Phys. Commun.} \textbf{\bibinfo{volume}{177}},
  \bibinfo{pages}{399} (\bibinfo{year}{2007}), \eprint{hep-ph/0612134}.

\bibitem[{\citenamefont{Das et~al.}(2012)\citenamefont{Das, Ellwanger, and
  Teixeira}}]{Das:2011dg}
\bibinfo{author}{\bibfnamefont{D.}~\bibnamefont{Das}},
  \bibinfo{author}{\bibfnamefont{U.}~\bibnamefont{Ellwanger}},
  \bibnamefont{and} \bibinfo{author}{\bibfnamefont{A.~M.}
  \bibnamefont{Teixeira}}, \bibinfo{journal}{Comput. Phys. Commun.}
  \textbf{\bibinfo{volume}{183}}, \bibinfo{pages}{774} (\bibinfo{year}{2012}),
  \eprint{1106.5633}.

\bibitem[{\citenamefont{Muhlleitner et~al.}(2005)\citenamefont{Muhlleitner,
  Djouadi, and Mambrini}}]{Muhlleitner:2003vg}
\bibinfo{author}{\bibfnamefont{M.}~\bibnamefont{Muhlleitner}},
  \bibinfo{author}{\bibfnamefont{A.}~\bibnamefont{Djouadi}}, \bibnamefont{and}
  \bibinfo{author}{\bibfnamefont{Y.}~\bibnamefont{Mambrini}},
  \bibinfo{journal}{Comput. Phys. Commun.} \textbf{\bibinfo{volume}{168}},
  \bibinfo{pages}{46} (\bibinfo{year}{2005}), \eprint{hep-ph/0311167}.

\bibitem[{\citenamefont{Kang et~al.}(2012)\citenamefont{Kang, Li, and
  Li}}]{Kang:2012sy}
\bibinfo{author}{\bibfnamefont{Z.}~\bibnamefont{Kang}},
  \bibinfo{author}{\bibfnamefont{J.}~\bibnamefont{Li}}, \bibnamefont{and}
  \bibinfo{author}{\bibfnamefont{T.}~\bibnamefont{Li}}, \bibinfo{journal}{JHEP}
  \textbf{\bibinfo{volume}{11}}, \bibinfo{pages}{024} (\bibinfo{year}{2012}),
  \eprint{1201.5305}.

\bibitem[{\citenamefont{Bernon et~al.}(2014)\citenamefont{Bernon, Dumont, and
  Kraml}}]{Bernon:2014vta}
\bibinfo{author}{\bibfnamefont{J.}~\bibnamefont{Bernon}},
  \bibinfo{author}{\bibfnamefont{B.}~\bibnamefont{Dumont}}, \bibnamefont{and}
  \bibinfo{author}{\bibfnamefont{S.}~\bibnamefont{Kraml}},
  \bibinfo{journal}{Phys. Rev.} \textbf{\bibinfo{volume}{D90}},
  \bibinfo{pages}{071301} (\bibinfo{year}{2014}), \eprint{1409.1588}.

\bibitem[{\citenamefont{Bernon and Dumont}(2015)}]{Bernon:2015hsa}
\bibinfo{author}{\bibfnamefont{J.}~\bibnamefont{Bernon}} \bibnamefont{and}
  \bibinfo{author}{\bibfnamefont{B.}~\bibnamefont{Dumont}},
  \bibinfo{journal}{Eur. Phys. J.} \textbf{\bibinfo{volume}{C75}},
  \bibinfo{pages}{440} (\bibinfo{year}{2015}), \eprint{1502.04138}.

\bibitem[{\citenamefont{Ginzburg and Krawczyk}(2005)}]{Ginzburg:2004vp}
\bibinfo{author}{\bibfnamefont{I.~F.} \bibnamefont{Ginzburg}} \bibnamefont{and}
  \bibinfo{author}{\bibfnamefont{M.}~\bibnamefont{Krawczyk}},
  \bibinfo{journal}{Phys. Rev.} \textbf{\bibinfo{volume}{D72}},
  \bibinfo{pages}{115013} (\bibinfo{year}{2005}), \eprint{hep-ph/0408011}.

\bibitem[{\citenamefont{Davidson and Haber}(2005)}]{Davidson:2005cw}
\bibinfo{author}{\bibfnamefont{S.}~\bibnamefont{Davidson}} \bibnamefont{and}
  \bibinfo{author}{\bibfnamefont{H.~E.} \bibnamefont{Haber}},
  \bibinfo{journal}{Phys. Rev.} \textbf{\bibinfo{volume}{D72}},
  \bibinfo{pages}{035004} (\bibinfo{year}{2005}), \bibinfo{note}{[Erratum:
  Phys. Rev.D72,099902(2005)]}, \eprint{hep-ph/0504050}.

\bibitem[{\citenamefont{Haber and O'Neil}(2006)}]{Haber:2006ue}
\bibinfo{author}{\bibfnamefont{H.~E.} \bibnamefont{Haber}} \bibnamefont{and}
  \bibinfo{author}{\bibfnamefont{D.}~\bibnamefont{O'Neil}},
  \bibinfo{journal}{Phys. Rev.} \textbf{\bibinfo{volume}{D74}},
  \bibinfo{pages}{015018} (\bibinfo{year}{2006}), \eprint{hep-ph/0602242}.

\bibitem[{\citenamefont{Ellwanger and
  Rodriguez-Vazquez}(2015)}]{Ellwanger:2015uaz}
\bibinfo{author}{\bibfnamefont{U.}~\bibnamefont{Ellwanger}} \bibnamefont{and}
  \bibinfo{author}{\bibfnamefont{M.}~\bibnamefont{Rodriguez-Vazquez}}
  (\bibinfo{year}{2015}), \eprint{1512.04281}.

\bibitem[{\citenamefont{Ade et~al.}(2015)}]{Ade:2015xua}
\bibinfo{author}{\bibfnamefont{P.~A.~R.} \bibnamefont{Ade}}
  \bibnamefont{et~al.} (\bibinfo{collaboration}{Planck})
  (\bibinfo{year}{2015}), \eprint{1502.01589}.

\bibitem[{\citenamefont{Belanger et~al.}(2005)\citenamefont{Belanger, Boudjema,
  Hugonie, Pukhov, and Semenov}}]{Belanger:2005kh}
\bibinfo{author}{\bibfnamefont{G.}~\bibnamefont{Belanger}},
  \bibinfo{author}{\bibfnamefont{F.}~\bibnamefont{Boudjema}},
  \bibinfo{author}{\bibfnamefont{C.}~\bibnamefont{Hugonie}},
  \bibinfo{author}{\bibfnamefont{A.}~\bibnamefont{Pukhov}}, \bibnamefont{and}
  \bibinfo{author}{\bibfnamefont{A.}~\bibnamefont{Semenov}},
  \bibinfo{journal}{JCAP} \textbf{\bibinfo{volume}{0509}}, \bibinfo{pages}{001}
  (\bibinfo{year}{2005}), \eprint{hep-ph/0505142}.

\bibitem[{\citenamefont{Belanger et~al.}(2014)\citenamefont{Belanger, Boudjema,
  Pukhov, and Semenov}}]{Belanger:2013oya}
\bibinfo{author}{\bibfnamefont{G.}~\bibnamefont{Belanger}},
  \bibinfo{author}{\bibfnamefont{F.}~\bibnamefont{Boudjema}},
  \bibinfo{author}{\bibfnamefont{A.}~\bibnamefont{Pukhov}}, \bibnamefont{and}
  \bibinfo{author}{\bibfnamefont{A.}~\bibnamefont{Semenov}},
  \bibinfo{journal}{Comput. Phys. Commun.} \textbf{\bibinfo{volume}{185}},
  \bibinfo{pages}{960} (\bibinfo{year}{2014}), \eprint{1305.0237}.

\bibitem[{\citenamefont{Akerib et~al.}(2015)}]{Akerib:2015rjg}
\bibinfo{author}{\bibfnamefont{D.~S.} \bibnamefont{Akerib}}
  \bibnamefont{et~al.} (\bibinfo{collaboration}{LUX}) (\bibinfo{year}{2015}),
  \eprint{1512.03506}.

\bibitem[{\citenamefont{Aad et~al.}(2014)}]{Aad:2014vma}
\bibinfo{author}{\bibfnamefont{G.}~\bibnamefont{Aad}} \bibnamefont{et~al.}
  (\bibinfo{collaboration}{ATLAS}), \bibinfo{journal}{JHEP}
  \textbf{\bibinfo{volume}{05}}, \bibinfo{pages}{071} (\bibinfo{year}{2014}),
  \eprint{1403.5294}.

\bibitem[{\citenamefont{Cheng et~al.}(2014)\citenamefont{Cheng, Li, Li, and
  Yan}}]{Cheng:2013fma}
\bibinfo{author}{\bibfnamefont{T.}~\bibnamefont{Cheng}},
  \bibinfo{author}{\bibfnamefont{J.}~\bibnamefont{Li}},
  \bibinfo{author}{\bibfnamefont{T.}~\bibnamefont{Li}}, \bibnamefont{and}
  \bibinfo{author}{\bibfnamefont{Q.-S.} \bibnamefont{Yan}},
  \bibinfo{journal}{Phys. Rev.} \textbf{\bibinfo{volume}{D89}},
  \bibinfo{pages}{015015} (\bibinfo{year}{2014}), \eprint{1304.3182}.

\bibitem[{\citenamefont{Guo et~al.}(2014)\citenamefont{Guo, Kang, Li, Li, and
  Liu}}]{Guo:2013asa}
\bibinfo{author}{\bibfnamefont{J.}~\bibnamefont{Guo}},
  \bibinfo{author}{\bibfnamefont{Z.}~\bibnamefont{Kang}},
  \bibinfo{author}{\bibfnamefont{J.}~\bibnamefont{Li}},
  \bibinfo{author}{\bibfnamefont{T.}~\bibnamefont{Li}}, \bibnamefont{and}
  \bibinfo{author}{\bibfnamefont{Y.}~\bibnamefont{Liu}},
  \bibinfo{journal}{JHEP} \textbf{\bibinfo{volume}{10}}, \bibinfo{pages}{164}
  (\bibinfo{year}{2014}), \eprint{1312.2821}.

\bibitem[{\citenamefont{Chatrchyan et~al.}(2014)}]{Chatrchyan:2013fea}
\bibinfo{author}{\bibfnamefont{S.}~\bibnamefont{Chatrchyan}}
  \bibnamefont{et~al.} (\bibinfo{collaboration}{CMS}), \bibinfo{journal}{JHEP}
  \textbf{\bibinfo{volume}{01}}, \bibinfo{pages}{163} (\bibinfo{year}{2014}),
  \bibinfo{note}{[Erratum: JHEP01,014(2015)]}, \eprint{1311.6736}.

\bibitem[{\citenamefont{Alwall et~al.}(2014)\citenamefont{Alwall, Frederix,
  Frixione, Hirschi, Maltoni, Mattelaer, Shao, Stelzer, Torrielli, and
  Zaro}}]{Alwall:2014hca}
\bibinfo{author}{\bibfnamefont{J.}~\bibnamefont{Alwall}},
  \bibinfo{author}{\bibfnamefont{R.}~\bibnamefont{Frederix}},
  \bibinfo{author}{\bibfnamefont{S.}~\bibnamefont{Frixione}},
  \bibinfo{author}{\bibfnamefont{V.}~\bibnamefont{Hirschi}},
  \bibinfo{author}{\bibfnamefont{F.}~\bibnamefont{Maltoni}},
  \bibinfo{author}{\bibfnamefont{O.}~\bibnamefont{Mattelaer}},
  \bibinfo{author}{\bibfnamefont{H.~S.} \bibnamefont{Shao}},
  \bibinfo{author}{\bibfnamefont{T.}~\bibnamefont{Stelzer}},
  \bibinfo{author}{\bibfnamefont{P.}~\bibnamefont{Torrielli}},
  \bibnamefont{and} \bibinfo{author}{\bibfnamefont{M.}~\bibnamefont{Zaro}},
  \bibinfo{journal}{JHEP} \textbf{\bibinfo{volume}{07}}, \bibinfo{pages}{079}
  (\bibinfo{year}{2014}), \eprint{1405.0301}.

\bibitem[{\citenamefont{Alloul et~al.}(2014)\citenamefont{Alloul, Christensen,
  Degrande, Duhr, and Fuks}}]{Alloul:2013bka}
\bibinfo{author}{\bibfnamefont{A.}~\bibnamefont{Alloul}},
  \bibinfo{author}{\bibfnamefont{N.~D.} \bibnamefont{Christensen}},
  \bibinfo{author}{\bibfnamefont{C.}~\bibnamefont{Degrande}},
  \bibinfo{author}{\bibfnamefont{C.}~\bibnamefont{Duhr}}, \bibnamefont{and}
  \bibinfo{author}{\bibfnamefont{B.}~\bibnamefont{Fuks}},
  \bibinfo{journal}{Comput. Phys. Commun.} \textbf{\bibinfo{volume}{185}},
  \bibinfo{pages}{2250} (\bibinfo{year}{2014}), \eprint{1310.1921}.

\bibitem[{\citenamefont{Duhr and Fuks}(2011)}]{Duhr:2011se}
\bibinfo{author}{\bibfnamefont{C.}~\bibnamefont{Duhr}} \bibnamefont{and}
  \bibinfo{author}{\bibfnamefont{B.}~\bibnamefont{Fuks}},
  \bibinfo{journal}{Comput. Phys. Commun.} \textbf{\bibinfo{volume}{182}},
  \bibinfo{pages}{2404} (\bibinfo{year}{2011}), \eprint{1102.4191}.

\bibitem[{\citenamefont{Degrande et~al.}(2012)\citenamefont{Degrande, Duhr,
  Fuks, Grellscheid, Mattelaer, and Reiter}}]{Degrande:2011ua}
\bibinfo{author}{\bibfnamefont{C.}~\bibnamefont{Degrande}},
  \bibinfo{author}{\bibfnamefont{C.}~\bibnamefont{Duhr}},
  \bibinfo{author}{\bibfnamefont{B.}~\bibnamefont{Fuks}},
  \bibinfo{author}{\bibfnamefont{D.}~\bibnamefont{Grellscheid}},
  \bibinfo{author}{\bibfnamefont{O.}~\bibnamefont{Mattelaer}},
  \bibnamefont{and} \bibinfo{author}{\bibfnamefont{T.}~\bibnamefont{Reiter}},
  \bibinfo{journal}{Comput. Phys. Commun.} \textbf{\bibinfo{volume}{183}},
  \bibinfo{pages}{1201} (\bibinfo{year}{2012}), \eprint{1108.2040}.

\bibitem[{\citenamefont{Christensen et~al.}(2011)\citenamefont{Christensen,
  de~Aquino, Degrande, Duhr, Fuks, Herquet, Maltoni, and
  Schumann}}]{Christensen:2009jx}
\bibinfo{author}{\bibfnamefont{N.~D.} \bibnamefont{Christensen}},
  \bibinfo{author}{\bibfnamefont{P.}~\bibnamefont{de~Aquino}},
  \bibinfo{author}{\bibfnamefont{C.}~\bibnamefont{Degrande}},
  \bibinfo{author}{\bibfnamefont{C.}~\bibnamefont{Duhr}},
  \bibinfo{author}{\bibfnamefont{B.}~\bibnamefont{Fuks}},
  \bibinfo{author}{\bibfnamefont{M.}~\bibnamefont{Herquet}},
  \bibinfo{author}{\bibfnamefont{F.}~\bibnamefont{Maltoni}}, \bibnamefont{and}
  \bibinfo{author}{\bibfnamefont{S.}~\bibnamefont{Schumann}},
  \bibinfo{journal}{Eur. Phys. J.} \textbf{\bibinfo{volume}{C71}},
  \bibinfo{pages}{1541} (\bibinfo{year}{2011}), \eprint{0906.2474}.

\bibitem[{\citenamefont{Skands et~al.}(2004)}]{Skands:2003cj}
\bibinfo{author}{\bibfnamefont{P.~Z.} \bibnamefont{Skands}}
  \bibnamefont{et~al.}, \bibinfo{journal}{JHEP} \textbf{\bibinfo{volume}{07}},
  \bibinfo{pages}{036} (\bibinfo{year}{2004}), \eprint{hep-ph/0311123}.

\bibitem[{\citenamefont{Allanach et~al.}(2009)}]{Allanach:2008qq}
\bibinfo{author}{\bibfnamefont{B.~C.} \bibnamefont{Allanach}}
  \bibnamefont{et~al.}, \bibinfo{journal}{Comput. Phys. Commun.}
  \textbf{\bibinfo{volume}{180}}, \bibinfo{pages}{8} (\bibinfo{year}{2009}),
  \eprint{0801.0045}.

\bibitem[{\citenamefont{Butterworth et~al.}(2010)}]{Butterworth:2010ym}
\bibinfo{author}{\bibfnamefont{J.~M.} \bibnamefont{Butterworth}}
  \bibnamefont{et~al.}, in \emph{\bibinfo{booktitle}{{Physics at TeV colliders.
  Proceedings, 6th Workshop, dedicated to Thomas Binoth, Les Houches, France,
  June 8-26, 2009}}} (\bibinfo{year}{2010}), \eprint{1003.1643}.

\bibitem[{\citenamefont{Sjostrand et~al.}(2006)\citenamefont{Sjostrand, Mrenna,
  and Skands}}]{Sjostrand:2006za}
\bibinfo{author}{\bibfnamefont{T.}~\bibnamefont{Sjostrand}},
  \bibinfo{author}{\bibfnamefont{S.}~\bibnamefont{Mrenna}}, \bibnamefont{and}
  \bibinfo{author}{\bibfnamefont{P.~Z.} \bibnamefont{Skands}},
  \bibinfo{journal}{JHEP} \textbf{\bibinfo{volume}{05}}, \bibinfo{pages}{026}
  (\bibinfo{year}{2006}), \eprint{hep-ph/0603175}.

\bibitem[{\citenamefont{Jadach et~al.}(1993)\citenamefont{Jadach, Was, Decker,
  and Kuhn}}]{Jadach:1993hs}
\bibinfo{author}{\bibfnamefont{S.}~\bibnamefont{Jadach}},
  \bibinfo{author}{\bibfnamefont{Z.}~\bibnamefont{Was}},
  \bibinfo{author}{\bibfnamefont{R.}~\bibnamefont{Decker}}, \bibnamefont{and}
  \bibinfo{author}{\bibfnamefont{J.~H.} \bibnamefont{Kuhn}},
  \bibinfo{journal}{Comput. Phys. Commun.} \textbf{\bibinfo{volume}{76}},
  \bibinfo{pages}{361} (\bibinfo{year}{1993}).

\bibitem[{\citenamefont{Davidson et~al.}(2012)\citenamefont{Davidson, Nanava,
  Przedzinski, Richter-Was, and Was}}]{Davidson:2010rw}
\bibinfo{author}{\bibfnamefont{N.}~\bibnamefont{Davidson}},
  \bibinfo{author}{\bibfnamefont{G.}~\bibnamefont{Nanava}},
  \bibinfo{author}{\bibfnamefont{T.}~\bibnamefont{Przedzinski}},
  \bibinfo{author}{\bibfnamefont{E.}~\bibnamefont{Richter-Was}},
  \bibnamefont{and} \bibinfo{author}{\bibfnamefont{Z.}~\bibnamefont{Was}},
  \bibinfo{journal}{Comput. Phys. Commun.} \textbf{\bibinfo{volume}{183}},
  \bibinfo{pages}{821} (\bibinfo{year}{2012}), \eprint{1002.0543}.

\bibitem[{\citenamefont{de~Favereau et~al.}(2014)\citenamefont{de~Favereau,
  Delaere, Demin, Giammanco, Lemaître, Mertens, and
  Selvaggi}}]{deFavereau:2013fsa}
\bibinfo{author}{\bibfnamefont{J.}~\bibnamefont{de~Favereau}},
  \bibinfo{author}{\bibfnamefont{C.}~\bibnamefont{Delaere}},
  \bibinfo{author}{\bibfnamefont{P.}~\bibnamefont{Demin}},
  \bibinfo{author}{\bibfnamefont{A.}~\bibnamefont{Giammanco}},
  \bibinfo{author}{\bibfnamefont{V.}~\bibnamefont{Lemaître}},
  \bibinfo{author}{\bibfnamefont{A.}~\bibnamefont{Mertens}}, \bibnamefont{and}
  \bibinfo{author}{\bibfnamefont{M.}~\bibnamefont{Selvaggi}}
  (\bibinfo{collaboration}{DELPHES 3}), \bibinfo{journal}{JHEP}
  \textbf{\bibinfo{volume}{02}}, \bibinfo{pages}{057} (\bibinfo{year}{2014}),
  \eprint{1307.6346}.

\bibitem[{\citenamefont{Cacciari et~al.}(2008)\citenamefont{Cacciari, Salam,
  and Soyez}}]{Cacciari:2008gp}
\bibinfo{author}{\bibfnamefont{M.}~\bibnamefont{Cacciari}},
  \bibinfo{author}{\bibfnamefont{G.~P.} \bibnamefont{Salam}}, \bibnamefont{and}
  \bibinfo{author}{\bibfnamefont{G.}~\bibnamefont{Soyez}},
  \bibinfo{journal}{JHEP} \textbf{\bibinfo{volume}{04}}, \bibinfo{pages}{063}
  (\bibinfo{year}{2008}), \eprint{0802.1189}.

\bibitem[{\citenamefont{Cacciari et~al.}(2012)\citenamefont{Cacciari, Salam,
  and Soyez}}]{Cacciari:2011ma}
\bibinfo{author}{\bibfnamefont{M.}~\bibnamefont{Cacciari}},
  \bibinfo{author}{\bibfnamefont{G.~P.} \bibnamefont{Salam}}, \bibnamefont{and}
  \bibinfo{author}{\bibfnamefont{G.}~\bibnamefont{Soyez}},
  \bibinfo{journal}{Eur. Phys. J.} \textbf{\bibinfo{volume}{C72}},
  \bibinfo{pages}{1896} (\bibinfo{year}{2012}), \eprint{1111.6097}.

\bibitem[{\citenamefont{Aad et~al.}(2012{\natexlab{b}})}]{Aad:2012mea}
\bibinfo{author}{\bibfnamefont{G.}~\bibnamefont{Aad}} \bibnamefont{et~al.}
  (\bibinfo{collaboration}{ATLAS}), \bibinfo{journal}{JHEP}
  \textbf{\bibinfo{volume}{09}}, \bibinfo{pages}{070}
  (\bibinfo{year}{2012}{\natexlab{b}}), \eprint{1206.5971}.

\bibitem[{\citenamefont{Li and Williams}(2015)}]{Li:2015sza}
\bibinfo{author}{\bibfnamefont{J.}~\bibnamefont{Li}} \bibnamefont{and}
  \bibinfo{author}{\bibfnamefont{A.~G.} \bibnamefont{Williams}}
  (\bibinfo{year}{2015}), \eprint{1508.05675}.

\bibitem[{\citenamefont{Aad et~al.}(2015{\natexlab{b}})}]{Aad:2014rga}
\bibinfo{author}{\bibfnamefont{G.}~\bibnamefont{Aad}} \bibnamefont{et~al.}
  (\bibinfo{collaboration}{ATLAS}), \bibinfo{journal}{Eur. Phys. J.}
  \textbf{\bibinfo{volume}{C75}}, \bibinfo{pages}{303}
  (\bibinfo{year}{2015}{\natexlab{b}}), \eprint{1412.7086}.

\bibitem[{\citenamefont{Kim}(2011)}]{Kim:2010uj}
\bibinfo{author}{\bibfnamefont{J.-H.} \bibnamefont{Kim}},
  \bibinfo{journal}{Phys. Rev.} \textbf{\bibinfo{volume}{D83}},
  \bibinfo{pages}{011502} (\bibinfo{year}{2011}), \eprint{1011.1493}.

\bibitem[{\citenamefont{Thaler and Van~Tilburg}(2011)}]{Thaler:2010tr}
\bibinfo{author}{\bibfnamefont{J.}~\bibnamefont{Thaler}} \bibnamefont{and}
  \bibinfo{author}{\bibfnamefont{K.}~\bibnamefont{Van~Tilburg}},
  \bibinfo{journal}{JHEP} \textbf{\bibinfo{volume}{03}}, \bibinfo{pages}{015}
  (\bibinfo{year}{2011}), \eprint{1011.2268}.

\bibitem[{\citenamefont{Debove et~al.}(2008)\citenamefont{Debove, Fuks, and
  Klasen}}]{Debove:2008nr}
\bibinfo{author}{\bibfnamefont{J.}~\bibnamefont{Debove}},
  \bibinfo{author}{\bibfnamefont{B.}~\bibnamefont{Fuks}}, \bibnamefont{and}
  \bibinfo{author}{\bibfnamefont{M.}~\bibnamefont{Klasen}},
  \bibinfo{journal}{Phys. Rev.} \textbf{\bibinfo{volume}{D78}},
  \bibinfo{pages}{074020} (\bibinfo{year}{2008}), \eprint{0804.0423}.

\bibitem[{\citenamefont{Debove et~al.}(2010)\citenamefont{Debove, Fuks, and
  Klasen}}]{Debove:2009ia}
\bibinfo{author}{\bibfnamefont{J.}~\bibnamefont{Debove}},
  \bibinfo{author}{\bibfnamefont{B.}~\bibnamefont{Fuks}}, \bibnamefont{and}
  \bibinfo{author}{\bibfnamefont{M.}~\bibnamefont{Klasen}},
  \bibinfo{journal}{Phys. Lett.} \textbf{\bibinfo{volume}{B688}},
  \bibinfo{pages}{208} (\bibinfo{year}{2010}), \eprint{0907.1105}.

\bibitem[{\citenamefont{Debove et~al.}(2011{\natexlab{a}})\citenamefont{Debove,
  Fuks, and Klasen}}]{Debove:2011xj}
\bibinfo{author}{\bibfnamefont{J.}~\bibnamefont{Debove}},
  \bibinfo{author}{\bibfnamefont{B.}~\bibnamefont{Fuks}}, \bibnamefont{and}
  \bibinfo{author}{\bibfnamefont{M.}~\bibnamefont{Klasen}},
  \bibinfo{journal}{Nucl. Phys.} \textbf{\bibinfo{volume}{B849}},
  \bibinfo{pages}{64} (\bibinfo{year}{2011}{\natexlab{a}}), \eprint{1102.4422}.

\bibitem[{\citenamefont{Debove et~al.}(2011{\natexlab{b}})\citenamefont{Debove,
  Fuks, and Klasen}}]{Debove:2010kf}
\bibinfo{author}{\bibfnamefont{J.}~\bibnamefont{Debove}},
  \bibinfo{author}{\bibfnamefont{B.}~\bibnamefont{Fuks}}, \bibnamefont{and}
  \bibinfo{author}{\bibfnamefont{M.}~\bibnamefont{Klasen}},
  \bibinfo{journal}{Nucl. Phys.} \textbf{\bibinfo{volume}{B842}},
  \bibinfo{pages}{51} (\bibinfo{year}{2011}{\natexlab{b}}), \eprint{1005.2909}.

\bibitem[{\citenamefont{Fuks et~al.}(2012)\citenamefont{Fuks, Klasen, Lamprea,
  and Rothering}}]{Fuks:2012qx}
\bibinfo{author}{\bibfnamefont{B.}~\bibnamefont{Fuks}},
  \bibinfo{author}{\bibfnamefont{M.}~\bibnamefont{Klasen}},
  \bibinfo{author}{\bibfnamefont{D.~R.} \bibnamefont{Lamprea}},
  \bibnamefont{and}
  \bibinfo{author}{\bibfnamefont{M.}~\bibnamefont{Rothering}},
  \bibinfo{journal}{JHEP} \textbf{\bibinfo{volume}{10}}, \bibinfo{pages}{081}
  (\bibinfo{year}{2012}), \eprint{1207.2159}.

\bibitem[{\citenamefont{Fuks et~al.}(2013)\citenamefont{Fuks, Klasen, Lamprea,
  and Rothering}}]{Fuks:2013vua}
\bibinfo{author}{\bibfnamefont{B.}~\bibnamefont{Fuks}},
  \bibinfo{author}{\bibfnamefont{M.}~\bibnamefont{Klasen}},
  \bibinfo{author}{\bibfnamefont{D.~R.} \bibnamefont{Lamprea}},
  \bibnamefont{and}
  \bibinfo{author}{\bibfnamefont{M.}~\bibnamefont{Rothering}},
  \bibinfo{journal}{Eur. Phys. J.} \textbf{\bibinfo{volume}{C73}},
  \bibinfo{pages}{2480} (\bibinfo{year}{2013}), \eprint{1304.0790}.

\bibitem[{\citenamefont{Beenakker et~al.}(1999)\citenamefont{Beenakker, Klasen,
  Kramer, Plehn, Spira, and Zerwas}}]{Beenakker:1999xh}
\bibinfo{author}{\bibfnamefont{W.}~\bibnamefont{Beenakker}},
  \bibinfo{author}{\bibfnamefont{M.}~\bibnamefont{Klasen}},
  \bibinfo{author}{\bibfnamefont{M.}~\bibnamefont{Kramer}},
  \bibinfo{author}{\bibfnamefont{T.}~\bibnamefont{Plehn}},
  \bibinfo{author}{\bibfnamefont{M.}~\bibnamefont{Spira}}, \bibnamefont{and}
  \bibinfo{author}{\bibfnamefont{P.~M.} \bibnamefont{Zerwas}},
  \bibinfo{journal}{Phys. Rev. Lett.} \textbf{\bibinfo{volume}{83}},
  \bibinfo{pages}{3780} (\bibinfo{year}{1999}), \bibinfo{note}{[Erratum: Phys.
  Rev. Lett.100,029901(2008)]}, \eprint{hep-ph/9906298}.

\bibitem[{\citenamefont{{The ATLAS collaboration}}(2015{\natexlab{b}})}]{xtt}
\bibinfo{author}{\bibnamefont{{The ATLAS collaboration}}},
  \bibinfo{journal}{ATLAS-CONF-2015-033}  (\bibinfo{year}{2015}{\natexlab{b}}).

\bibitem[{\citenamefont{Campbell et~al.}(2011)\citenamefont{Campbell, Ellis,
  and Williams}}]{Campbell:2011bn}
\bibinfo{author}{\bibfnamefont{J.~M.} \bibnamefont{Campbell}},
  \bibinfo{author}{\bibfnamefont{R.~K.} \bibnamefont{Ellis}}, \bibnamefont{and}
  \bibinfo{author}{\bibfnamefont{C.}~\bibnamefont{Williams}},
  \bibinfo{journal}{JHEP} \textbf{\bibinfo{volume}{07}}, \bibinfo{pages}{018}
  (\bibinfo{year}{2011}), \eprint{1105.0020}.

\bibitem[{\citenamefont{Lester and Summers}(1999)}]{Lester:1999tx}
\bibinfo{author}{\bibfnamefont{C.~G.} \bibnamefont{Lester}} \bibnamefont{and}
  \bibinfo{author}{\bibfnamefont{D.~J.} \bibnamefont{Summers}},
  \bibinfo{journal}{Phys. Lett.} \textbf{\bibinfo{volume}{B463}},
  \bibinfo{pages}{99} (\bibinfo{year}{1999}), \eprint{hep-ph/9906349}.

\bibitem[{\citenamefont{Cheng and Han}(2008)}]{Cheng:2008hk}
\bibinfo{author}{\bibfnamefont{H.-C.} \bibnamefont{Cheng}} \bibnamefont{and}
  \bibinfo{author}{\bibfnamefont{Z.}~\bibnamefont{Han}},
  \bibinfo{journal}{JHEP} \textbf{\bibinfo{volume}{12}}, \bibinfo{pages}{063}
  (\bibinfo{year}{2008}), \eprint{0810.5178}.

\end{thebibliography}

\end{document}